\documentclass[fleqn,usenatbib]{mnras}

\usepackage{newtxtext,newtxmath}

\usepackage[T1]{fontenc}

\DeclareRobustCommand{\VAN}[3]{#2}
\let\VANthebibliography\thebibliography
\def\thebibliography{\DeclareRobustCommand{\VAN}[3]{##3}\VANthebibliography}

\usepackage{graphicx}	
\usepackage{amsmath}	

\usepackage{cleveref}
\usepackage{siunitx}
\usepackage{longtable}

\crefformat{section}{\S#2#1#3} 
\crefformat{subsection}{\S#2#1#3}
\crefformat{subsubsection}{\S#2#1#3}

\usepackage{lipsum}
\usepackage[english]{babel}
\usepackage{color}
\usepackage[dvipsnames]{xcolor}

\usepackage{romannum}

\title[HM Cancri]{Two decades of optical timing of the shortest-period binary star system HM Cancri}
\author[J.\ Munday et al.]{James Munday,$^{1}$\thanks{Email: james.munday98@gmail.com}
T. R. Marsh,$^{1}$
Mark Hollands,$^{2}$
Ingrid Pelisoli,$^{1}$
Danny Steeghs,$^{1}$
Pasi Hakala,$^{3}$
\newauthor
Elm\'{e} Breedt,$^{4}$
Alex Brown,$^{2}$
V. S.\ Dhillon,$^{2,5}$
Martin J. Dyer,$^{2}$
Matthew Green,$^{6}$
Paul Kerry,$^{2}$
\newauthor
S.P.\ Littlefair,$^{2}$
Steven G. Parsons,$^{2}$
Dave Sahman,$^{2}$
Sorawit Somjit,$^{7}$
Boonchoo Sukaum,$^{7}$
James Wild$^{2}$ \\
$^{1}$Department of Physics, Gibbet Hill Road, University of Warwick, Coventry CV4 7AL, United Kingdom\\
$^{2}$Department of Physics and Astronomy, University of Sheffield, Sheffield, S3 7RH, UK\\
$^{3}$Finnish Centre for Astronomy with ESO (FINCA), Quantum, University of Turku, FI-20014, Turku, Finland\\
$^{4}$Institute of Astronomy, University of Cambridge, Madingley Road, Cambridge CB3 0HA, UK\\
$^{5}$Instituto de Astrof\'isica de Canarias, E-38205 La Laguna, Spain\\
$^{6}$Department of Astrophysics, School of Physics and Astronomy, Tel Aviv University, Tel Aviv 6997801, Israel\\
$^{7}$National Astronomical Research Institute of Thailand, 191 Siriphanich Building, Huay Kaew Road, Chiang Mai 50200, Thailand
}

\date{Accepted 2022 November 16. Received 2022 November 16; in original form 2022 September 1}

\pubyear{2022}

\begin{document}
\label{firstpage}
\pagerange{\pageref{firstpage}--\pageref{lastpage}}
\maketitle

\begin{abstract}
The shortest-period binary star system known to date, RX J0806.3+1527 (HM Cancri), has now been observed in the optical for more than two decades. Although it is thought to be a double degenerate binary undergoing mass transfer, an early surprise was that its orbital frequency, $f_0$, is currently increasing as the result of gravitational wave radiation. This is unusual since it was expected that the mass donor was degenerate and would expand on mass loss, leading to a decreasing $f_0$. We exploit two decades of high-speed photometry to precisely quantify the trajectory of HM~Cancri, allowing us to find that $\ddot f_0$ is negative, where $\ddot f_0~=~(-5.38\pm2.10)\times10^{-27}$~Hz\,s$^{-2}$. Coupled with our positive frequency derivative, we show that mass transfer is counteracting gravitational-wave dominated orbital decay and that HM~Cnc will turn around within $2100\pm800$\,yrs from now. We present Hubble Space Telescope ultra-violet spectra which display Lyman-$\alpha$ absorption, indicative of the presence of hydrogen accreted from the donor star. We use these pieces of information to explore a grid of permitted donor and accretor masses with the Modules for Experiments in Stellar Astrophysics suite, finding models in good accordance with many of the observed properties for a cool and initially hydrogen-rich extremely-low-mass white dwarf ($\approx$0.17\,\(\textup{M}_\odot\)) coupled with a high accretor mass white dwarf ($\approx$1.0\,\(\textup{M}_\odot\)). Our measurements and models affirm that HM~Cancri is still one of the brightest verification binaries for the Laser Interferometer Space Antenna spacecraft.
\end{abstract}




\section{Introduction}
There exists a large population of double white dwarf (DWD) binary star systems in the galaxy \citep{NelemansI2001populationSynthesisOfWDs, NelemansII2001populationSynthesisOfWDsAMCVn} which can form through a series of mass-transfer and common envelope events \citep[e.g.][]{Hurley2002BinaryEvolution}. Gravitational wave radiation drives the loss of orbital angular momentum and causes many binary orbits to decay within a Hubble time, driving them to compact (orbital period of $\approx$5--65\,min) configurations. Ultimately, these systems initiate mass transfer, which has a dramatic effect upon their evolution, making DWD binaries the likely progenitors of type \Romannum{1}a/.\Romannum{1}a supernovae \citep[][]{Maoz2014Type1aProgenitors} and maybe the progenitors of R~CrB stars \citep[][]{Webbink1984DWDprogenitorsRCrB}, hot subdwarf stars \citep[][]{Zhang2012FormationOfHeRichHotSubdwarfs,Bertolami2022SubdwarfFromCOWDplusHEWD} and AM~CVn systems \citep[][]{NelemansII2001populationSynthesisOfWDsAMCVn}. In contrast to the typical dynamics of mass transfer between non-degenerate stars, the smaller mass, larger radius white dwarf (WD) is first to overflow its Roche lobe as the donor star, with the larger mass, smaller radius WD the accretor in a DWD binary. Mass transfer from one star to a higher mass companion acts to expand the orbit, counteracting the shrinkage caused by gravitational wave losses with angular momentum transfer playing an important role in the fate of these systems \citep{Marsh2004WDMassTransfer}.

RX J0806.3+1527 \citep[HM Cancri, henceforth HM~Cnc, ][and also independently discovered by \citealt{Israel2002HMCncDiscovery}]{RamsayHakalaCropper2002HMCncBinarydiscoveryMaybe} is the shortest-period binary star system known, having a present day orbital period of 5.36\,min, equivalent to an orbital frequency of 3.11$\,$mHz. There have been multiple theories suggested to explain the extremely short period of HM~Cnc, namely the unipolar inductor model \citep[][]{Wu2002UImodel, DallOsso2007UIHMCnc}; the intermediate polar model \citep[][]{Norton2004IPmodelHMCnc}; the AM~CVn model \citep[e.g.][]{NelemansII2001populationSynthesisOfWDsAMCVn}. In brief, the unipolar inductor model is similar to the Jupiter-Io system with the binary being detached, the intermediate polar model suggests that the optical period actually represents the spins of magnetic WDs and that the binary's orbital period is much longer, and the AM~CVn model involves a compact WD binary mass-transferring helium rich material. In the case of HM~Cnc, a DWD binary is required to reach the observed period for an AM~CVn without a merger already taking place. The studies of \citet[][]{Roelofs2010HMCncMassRatio} and \citet[][]{Mason2010HMcncSpectraVLT} helped to reduce the options. They find clear radial-velocity variations on the \SI{5.36}{\min} optical period, difficult to explain using the intermediate polar model, and a strong indication of an accretion stream via the presence of He$\,${\sc{ii}} emission lines in the spectrum, which is not predicted by the unipolar inductor model. \citet[][]{Strohmayer2008HMCncHighResXray} fail to detect any signature of metals in the X-ray spectrum, favouring a DWD AM~CVn, and a DWD binary is also a natural explanation for the relative phase difference between the peak of X-ray and optical flux \citep[e.g.][]{Barros2007HMCnc}. Here, the peak in X-ray flux arrives slightly later than the optical, which arises from the trajectory of the accretion stream directly impacting the surface of the accretor. It would thus seem that only the DWD AM~CVn theory can be used to explain all of the present observations.

Follow-up observations of HM~Cnc in the X-ray and optical domains have revealed an increase in orbital frequency, $f$, due to the loss of orbital angular momentum through gravitational wave radiation \citep{Hakala2003HMCncBinary, Israel2004HMCncPulse, Hakala2004HMCncmore, Strohmayer2005timings, Barros2007HMCnc, Esposito2014HMCncSWIFT, Strohmayer2021HMCnc}. The increase is so much so that HM~Cnc will be one of the brightest ``verification binaries'' to optimise the performance of the Laser Interferometer Space Antenna \citep[LISA,][]{LISA} spacecraft, given its low-frequency gravitational wave emission, strong signal strength \citep{Kupfer2018LISAverificationBinaries, AmaroSeoane2022LISAastrophysics} and the precise timing measurements that have been used to quantify its orbital decay. Similarly, HM~Cnc is planned to be used as the primary reference source for the TianQin spacecraft \citep[][]{TianQinProposal2016,TianQinOptimisingOrbits}. While it took two years to detect an increase of HM~Cnc's orbital frequency \citep{Hakala2003HMCncBinary}, continuous timing measurements over a much longer baseline can prove insightful to determine the second derivative, $\ddot f$, which is an indicator of the trajectory of inspiral.

Using X-ray data, \citet[][]{Strohmayer2021HMCnc} measured the second derivative of the orbital frequency of HM~Cnc, finding $\ddot f = (-8.95\pm1.40) \times 10^{-27}\,$Hz$\,$s$^{-2}$. This was the first significant detection of $\ddot{f}$ for any AM~CVn system. The combination of $\dot f$ and $\ddot f$ can be used to predict the timing of maximum orbital frequency of the binary, which  \citet{Strohmayer2021HMCnc} predicts to occur in $1260\pm200\,$yrs from now. In addition, \citet{Strohmayer2021HMCnc} was able to estimate a mass transfer rate based on the accretion luminosity of the source from X-ray observations. HM~Cnc is unfortunately too dim ($G_{\text{mag}} = 20.9$) for a distance constraint with Gaia~DR3 \citep{GaiaMissionPaper}, though assuming a distance of 5$\,$kpc and donor and accretor masses of 0.27$\,$\(\textup{M}_\odot\) and 0.55$\,$\(\textup{M}_\odot\), \citet{Strohmayer2021HMCnc} calculated an accretion rate $\dot M = -1.6\times10^{-8}\,$\(\textup{M}_\odot\)$\,$yr$^{-1}$. However, the calculated mass transfer rate is strongly dependent on the input star masses and the distance to the source. A distance measurement and a precise measurement of $\ddot f$ are therefore extremely useful to constrain the binary's evolution with sets of donor and companion masses.

Our study is set out as follows. \cref{sec:Observations} discusses our photometric and spectroscopic observations of HM~Cnc and our attempts to derive a distance measurement. In \cref{sec:Method} we present system timing measurements and derive the orbital ephemeris. We then explore the extent to which the AM~CVn channel can match our orbital ephemeris with binary evolution simulations in \cref{sec:EvolutionaryModellingMESA}.

\section{Observations}
\label{sec:Observations}
\subsection{Photometry}
\label{subsec:photometry}
Time-series photometry was obtained using the high-speed cameras ULTRACAM \citep{ULTRACAM2007}, ULTRASPEC \citep{ULTRASPEC2014} and HiPERCAM \citep{HiPERCAM2016, Hipercam2021Paper}; ULTRACAM and HiPERCAM observe simultaneously in three or five bands, respectively. Data from HiPERCAM were obtained while mounted on the 4.2m William Herschel Telescope (WHT) and the 10.4m Gran Telescopio Canarias (GTC). We observed with ULTRACAM on the 8.2m Very Large Telescope (VLT), the 3.58m ESO New Technology Telescope (NTT) and the WHT, and we observed with ULTRASPEC mounted on the the 2.4m Thai National Observatory (TNO). Exposure times were varied across the nights to reflect the seeing conditions, the aperture of the telescope and the filters used at the time. A full observing log giving dates, conditions, exposure times and the filters used for each set of observations is given in Appendix~\ref{tab:observingLog}.
\begin{figure}
    \centering
    \includegraphics[trim={0.5cm 0.5cm 0.5cm 0.5cm},clip, width=\columnwidth, keepaspectratio]{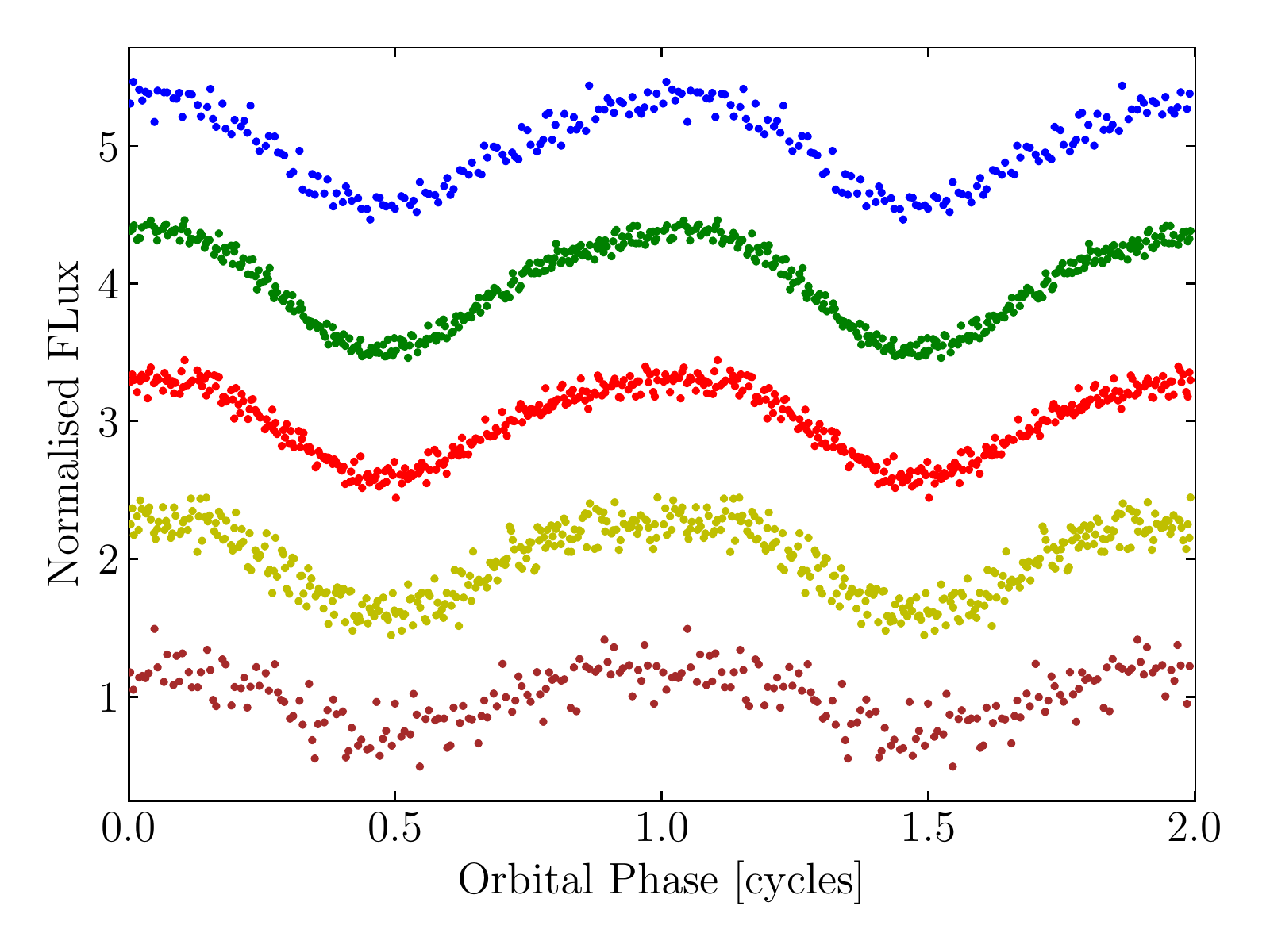}
    \caption{Phase-folded HM~Cnc light curves from HiPERCAM on the GTC. All filters are displayed. From top to bottom, the super SDSS $u_s$, $g_s$, $r_s$, $i_s$ and $z_s$. The non-sinusoidal nature of the light curves are noticeable around maximum flux, consequential of the flux contributed by the direct-impact spot. Each band is displaced by an artificial offset for clarity. The peak flux arrives first in the $z_s$ band and last in the $u_s$.}
    \label{fig:LCs}
\end{figure}
All of the data were bias-subtracted and flat-fielded. An additional dark current subtraction was applied to all data from ULTRACAM only as the instrument operates at a temperature high enough for there to be significant dark current. For HiPERCAM, corrections for CCD fringing were performed in the $z_s$ band only using pre-obtained fringe maps for the instrument.

We also accessed VLT FORS1 \citep[][]{VLTFors1998}, ESO 3.6m Telescope EFOSC2 \citep[][]{ESOEFOSC2} and Telescopio Nazionale Galileo (TNG) DoLoReS archival data that were presented in \citet[][]{Israel2002HMCncDiscovery} prior to our first observation. Extra archival data of HM~Cnc were obtained from the TNG DoLoReS instrument from dates following their study. Though the study of \citet[][]{Israel2002HMCncDiscovery} used some of these data before, no timing solutions have been reported, so we use the archival data in our study to take advantage of the longest observational baseline possible\footnote{All data presented in \citet[][]{Israel2002HMCncDiscovery} had calibration files and were bias-corrected and flat-fielded with custom scripts. Unfortunately in some cases either windowing setups were not specified or calibration files were not able to be recovered from the TNG archive, meaning that some data extraction was performed without bias or flat-field correction. Whenever HM~Cnc overlapped an artefact, or, the image showed a noticeably non-uniform local field, the data from this night were discarded.}. Data obtained with the 2.56m Nordic Optical Telescope (NOT) that was presented in \citet[][]{RamsayHakalaCropper2002HMCncBinarydiscoveryMaybe, Hakala2003HMCncBinary,Hakala2004HMCncmore} and with the 2.5m Isaac Newton Telescope (INT) from the RATS survey \citep[][]{RATShmCnc2005} were utilised in this study too. The extracted differential photometry obtained in prior studies using data from the NOT were used, whereas raw files were handled from every other source.

Photometry from each instrument, besides data obtained with the NOT, was then extracted using the HiPERCAM pipeline with differential aperture photometry and the application of \citet{OptimalPhotometry1998Naylor}'s optimal photometry algorithm when the target aperture did not need to be linked to a comparison star; otherwise regular aperture photometry was performed. To tackle poorer observing conditions, a variable aperture radius was applied, which was chosen to be a fixed multiple (1.6--1.8$\times$) of the full-width at half-maximum of the stars for each individual frame. A non-variable comparison star (Gaia DR3 ID~654873149889859968) was used for all observations when possible, or another (Gaia DR3 ID 654879025405159040) if the main comparison star saturated or coincided with detector defects.

The observation times were corrected to the Barycentric Julian Date (BJD) Barycentric Dynamical Time (TDB) centred on the mid-point of the exposure time. Poor data (e.g. affected by clouds) were rejected by visual inspection with HiPERCAM pipeline routines. We then derived a differential atmospheric extinction term for each filter to correct for the difference in colour between the comparison star and HM~Cnc. This was performed by considering the dependence of airmass on the differential flux, which was subtracted to effectively convert to a flux at zenith.

\subsection{Proper Motion}
\label{subsec:ProperMotion}
As of Gaia DR3, there is no measurement of the proper motion of HM~Cnc, so we searched for it using our data. To do so, we added world coordinate system information to each FITS file by plate solving each exposure using the backend of Astrometry.net \citep[][]{Astrometrynet2010}. We utilised a custom Gaia DR3 reference star catalogue for the local field containing stars with a right ascension~($\alpha$) and declination~($\delta$) error of less than 0.1$\,$arcseconds. The catalogue provided the location of stars at the mid-Gaia DR3 epoch. We selected data from the $V$, $r$ and $i$ bands only to reduce the impact of atmospheric refraction upon the position of HM~Cnc relative to the redder stars in the field, though atmospheric refraction still limits our position accuracy. A reliable coordinate calibration was possible for all frames with approximately 20 or more reference stars. Then, we converted the central (pixel) coordinate and error of the centroid for each aperture into a physical $\alpha$/$\delta$ coordinate. All measurements of $\alpha$/$\delta$ from each frame were then combined with a variance weighted mean, giving a measured $\alpha$/$\delta$ for HM~Cnc on each night.

From our determined positions (as plotted in Appendix~\ref{appendix:propermotionGraphs}), we conservatively take the range of the data to find upper limits of $|\dot\alpha|<9.1\,$mas\,yr$^{-1}$ and $|\dot\delta|<9.2\,$mas\,yr$^{-1}$. Our best-fit indicates proper motions of $\dot\alpha=2.0\pm1.1$~mas\,yr$^{-1}$ and $\dot\delta=-3.3\pm1.1$~mas\,yr$^{-1}$, having a reduced $\chi^2$ equal to one. By taking the maximum proper motion, we can derive a loose minimum distance constraint.

The proper motion is inversely proportional to the distance. Using an approximate mean transverse velocity ($V_T$) of 30$\,$km\,s$^{-1}$ \citep[][]{Cheng2019CoolingAndTransverseVel} for a WD in the galactic disc and a maximum absolute proper motion of $\dot\mu~=~\sqrt{|\dot\alpha|^2 + |\dot\delta|^2 \cos^2\left(\delta\right)}~\approx~12.7\,$mas\,yr$^{-1}$, we find a minimum distance of $D=500\,$pc. Our best-fit proper motion with $V_T=30\,$km\,s$^{-1}$ gives $D=(1.7\pm0.5)\,$kpc. This result is consistent with a distance limit of $D>1.1\,$kpc if X-rays are responsible for the heating of the donor, but goes against the limit set if optical/ultra-violet irradiation is responsible, being $d>4.2\,$kpc \citep[][]{Barros2007HMCnc}. Our best-fitting distance measurement is consistent with the expectation of $D\approx2$\,kpc predicted by \citet[][]{Reinsch2007HMCncFluxes}, although an absolute distance measurement is still relatively unconstrained by our assumption of a disc-like transverse velocity. Halo velocities are often larger \citep[e.g.][]{Pauli20063Dkinematics,Kim2020HaloVelocities} whereby assuming $V_T=200\,$km\,s$^{-1}$ we obtain a lower limit of $D=3300$\,pc and a best-fitting distance of $D=(11\pm3)$\,kpc. We hope that future Gaia data releases will reveal a more precise proper motion and perhaps a parallax measurement, though the $G_{\textrm{mag}}=20.9$ target is at the limit of Gaia's capabilities.

\subsection{Luminosity}
\label{subsec:Luminosity}
If the mass transfer rate of HM~Cnc changes as the orbit shrinks, so will the accretion luminosity. We used the ULTRACAM data using the prime and Super SDSS sets of filters to see if the brightness of HM~Cnc has changed over the years. We measured HM~Cnc's magnitude relative to comparison star Gaia~DR3~ID 654873149889859968 using the values of the constant offset $d$ (see \cref{subsec:FittingIndivRuns}) from equation~(\ref{eqn:WaveFit}). A difference in colour between HM~Cnc and the much redder comparison star was accounted for by subtracting filter-dependent colour coefficients as a function of airmass from the magnitudes obtained, as in \cref{subsec:photometry}. The measured changes in magnitude are presented in Table~\ref{tab:luminosity}. Observations with the Super SDSS filters have a total observing span of 4 years, while the SDSS prime observations span 13 years in $u'$ and $g'$, and about 7 years in $r'$ and $i'$, hence the change in magnitude measured with the SDSS prime filters have smaller uncertainties.

For the most part, our measurements favour a small brightening of HM~Cnc, but all measurements are within $3\sigma$ of no change at all, and so there has been no significant change overall.
\begin{table}
    \caption{Our measured change in magnitude of HM~Cnc in each filter set. HiPERCAM data is excluded from the brightness analysis due to an incompatible comparison star with the ULTRACAM data. This left one single night with the $r_s$ filter, such that no change in magnitude is reported.}
    \centering
    \begin{tabular}{c c c c}
        \hline
        Filter & $\Delta$mag (mag\,yr$^{-1}$) & Filter & $\Delta$mag (mag\,yr$^{-1}$) \\
        \hline
         u$_s$ & $-0.0149 \pm 0.0111$ & u' & $-0.0017 \pm 0.0036$\\
         g$_s$ & $-0.0177 \pm 0.0080$ & g' & $-0.0013 \pm 0.0029$\\
         i$_s$ & $-0.0686 \pm 0.0406$ & r'  & $-0.0113 \pm 0.0041$\\
         &&i'  & $\phantom{-}0.0080 \pm 0.0106$\\
         \hline
    \end{tabular}
    \label{tab:luminosity}
\end{table}

\subsection{Spectroscopy}
\label{subsec:spectroscopy}
\begin{figure}
    \centering
    \includegraphics[trim={0.5cm 0.5cm 0.5cm 0.25cm},clip,width=\columnwidth, keepaspectratio]{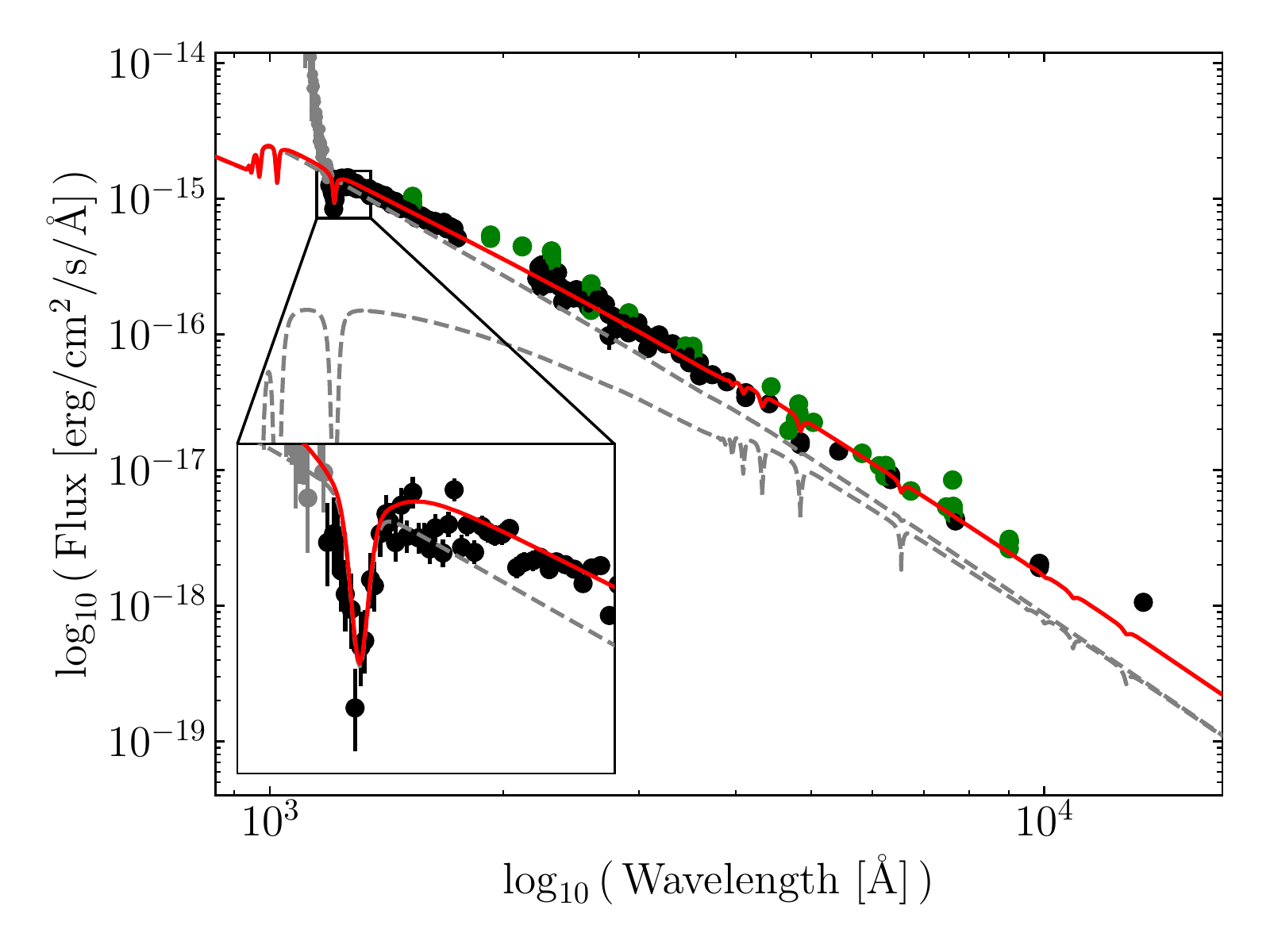}
    \caption{The HST stacked spectrum of HM~Cnc (black points) with our best spectral fit overplotted in red. Included in the figure is a magnified inset of the Lyman-$\alpha$ absorption feature. In every individual spectrum and at all orbital phases, Lyman-$\alpha$ was recognisable. Other hydrogen absorption lines are not directly visible in the spectrum and blend with helium emission lines \citep[][]{Roelofs2010HMCncMassRatio}, which are not modelled. In dashed grey are the contribution of flux from each star, with the donor contributing less flux. Grey circular points are from the wing of the far-ultraviolet spectrum that were not included in the fit. Green circular points are photometric fluxes from survey data obtained through CDS\protect\footnotemark, which were dereddened.}
    \label{fig:spectrum}
\end{figure}

We utilised the Hubble Space Telescope (HST) Advanced Camera for Surveys (ACS) to obtain slitless spectroscopy of HM~Cnc in 2007. We observed with the High Resolution Channel (HRC) with the PR200L prism (R$=300$--30) and with the Solar Blind Channel (SBC) with the PR110L prism (R$=60$--10), with spectral resolution decreasing as the wavelength increases due to a very non-linear prism dispersion. All spectra were individually reduced and flux calibrated with the HST aXe software and were then stacked\footnote{It was not possible to use the typical HST ``drizzle'' scheme to combine spectra due to the non-linear dispersion solution. Cosmic rays were flagged by the HST ACS pipeline and were subsequently removed.}. The resultant spectrum was de-reddened using the \textsc{extinction} python package following the treatment of \citet[][]{Fitzpatrick99}, with A$_V=0.10$\,mag \citep[][]{Esposito2014HMCncSWIFT} and R$_V=3.1$. The spectrum obtained is displayed in Fig.~\ref{fig:spectrum}. Some data at both ends of the spectra for each setup were deemed unreliable in the flux calibration due to a degraded resolution, creating a `winged' effect that is typical in the ACS slitless spectroscopy setup and discussed in detail in the instrument manual. The far ultra-violet end of the impacted regions were masked in our investigation and are shown as well in Fig.~\ref{fig:spectrum}.

\footnotetext{\text{https://cds.u-strasbg.fr/}}
Prior analyses of optical spectra have indirectly suggested that hydrogen is present due to blending of odd terms of the He\,{\sc{ii}} Pickering series \citep[][]{Norton2004IPmodelHMCnc, Reinsch2007HMCncFluxes}. Our stacked spectrum reveals a clear Lyman-$\alpha$ absorption feature in the ultra-violet, which is apparent in all spectra. This is the first direct evidence of hydrogen in the spectrum of HM~Cnc and signifies that hydrogen is present in the accretion stream and thus the envelope of the donor. HM~Cnc is the only known ultra-compact AM~CVn to show any direct trace of hydrogen, with the possible exception of CP~Eri \citep[][]{Sion2006CPEriHydrogen}.

Fitting to the observed spectral energy distribution allows us to obtain an improved temperature measurement of the accretor and to predict a hydrogen envelope fraction. The total mass accreted leading up to present day is expected to be approximately 0.01\,\(\textup{M}_\odot\) with the majority of mass transferred through direct impact accretion (see \cref{sec:EvolutionaryModellingMESA}), such that the surface composition of both stars will be near identical. To constrain these properties from the spectrum, we used the Koester WD atmosphere models \citep[][]{Koester2010WDmodels}. These allow arbitrary atmospheric compositions including hydrogen-helium mixtures to be used in both the atmospheric structure calculation and the spectral synthesis. We performed a least-squares fit to the data including both stellar components, with model inputs being the effective temperature of the donor and accretor, the relative scaling of the donor to the accretor (i.e. $R_{d}^2/R_{a}^2$, with $R$ the radius of the respective star and subscripts $d$ and $a$ the donor and the accretor stars) and a hydrogen-helium abundance which is identical for both stars with no metallicity. Inclusion of a scaling parameter is necessary because, with no firm distance constraint and only one spectral line in the low-resolution spectrum, it is not possible to constrain the surface gravity of either component, and so all calculations were performed at $\log g=8$. At each step in the least-squares analysis, the atmospheric structures and synthetic spectra of both components were calculated and summed together with the inclusion of the model scaling term. This combined spectrum was then convolved with a Gaussian function to match the instrumental resolution at Lyman-$\alpha$ ($\text{R}=300$). Finally, the model spectrum was scaled from a radiated flux, $F$, to a flux observed at Earth, $f_{\textrm{obs}}$, where
\begin{equation}
    f_{\textrm{obs}} = \frac{\pi R_a^2}{D^2} \left(F_a + F_d \frac{R_{d}^2}{R_{a}^2} \right)
\end{equation}
and $\pi R_a^2 / D^2 = 4.37\times 10^{-26}$ is our determined scaling constant.

The derived best-fitting solution is also displayed in Fig.~\ref{fig:spectrum}. From this model, we find a number abundance ratio of $\log(\textrm{H/He})=-1.64\pm0.05$\,dex ($\textrm{H/He}=2.29\pm0.26\%$), the hotter star (accretor) temperature to be $42\,500\pm800$\,K and the cooler star (donor) $18\,600\pm300$\,K. These quoted errors are purely statistical from the least-squares minimisation and do not reflect the errors in both our choice of reddening coefficients and the flux calibration when combining two instrumental setups, which are such that the cooler component remains largely unconstrained. Furthermore, it is the cooler star that contributes most significantly towards the redder wavelengths, where our spectrum is sparsely sampled and under-fits the observations. The relative contribution from the dimmer donor is $R_{d}^2$/$R_{a}^2=2.38\pm0.31$, however the radii themselves are unknown with little constraint on the masses of the two stars. If we consider R$_{a}=0.01$\,\(\textup{R}_\odot\) with our minimised scaling constant to convert from the radiated flux to the flux at Earth, we find $D=1.9$\,kpc. Doubling the accretor radius would double the inferred distance.

Although difficult at low resolution, other spectral lines were searched for in the stacked spectrum of all exposures and with stacked observations as a function of orbital phase. We were not able to detect any, besides the prominent Lyman-$\alpha$ absorption line. 

\section{Timing Solutions}
\label{sec:Method}
\subsection{Fitting individual observing runs}
\label{subsec:FittingIndivRuns}
\begin{figure*}
    \centering
    \includegraphics[trim={0.5cm 0 0.5cm 0},clip, width=\columnwidth, keepaspectratio]{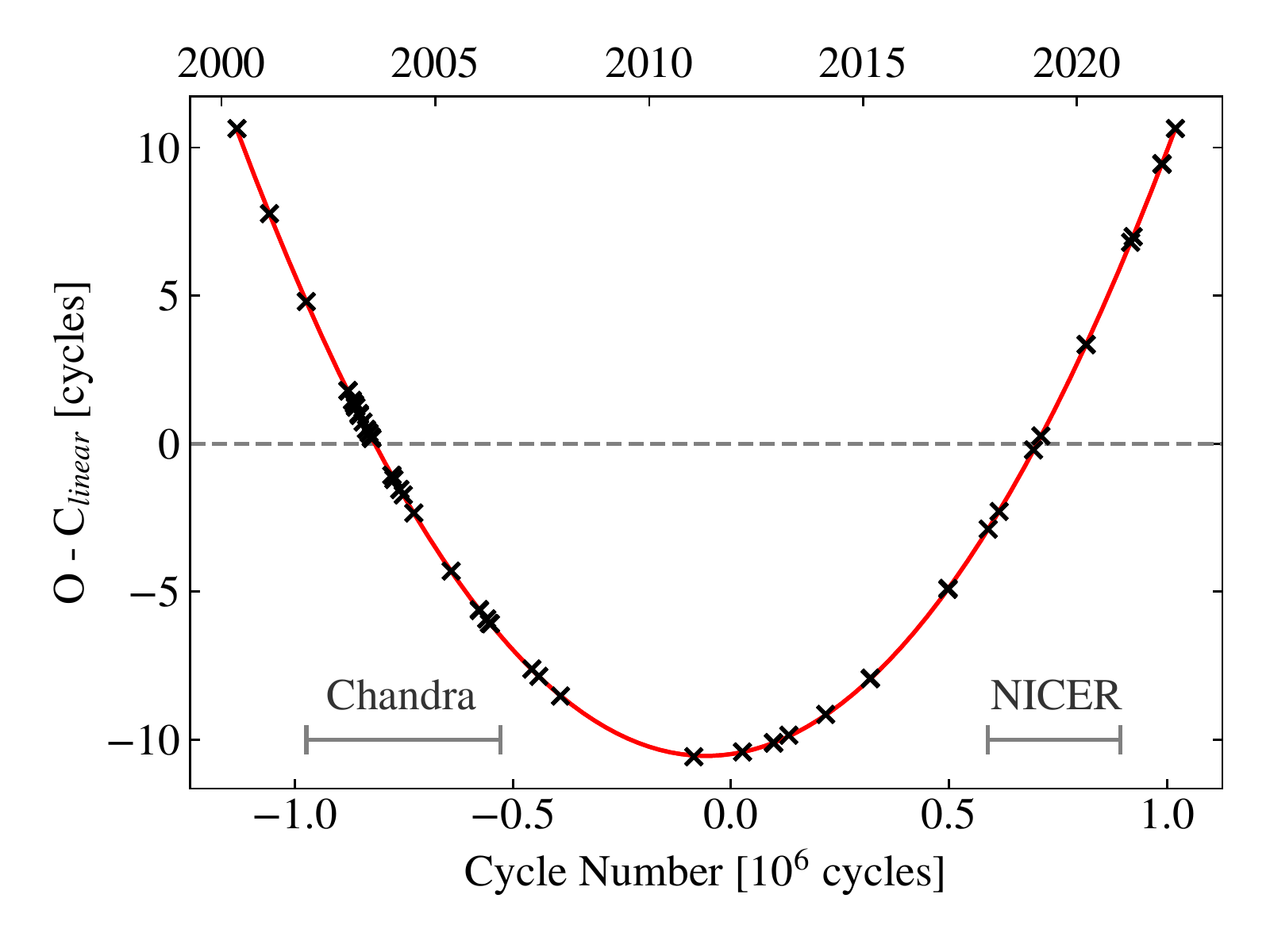}
    \hfill
    \includegraphics[trim={0.4cm 0 0.5cm 0},clip, width=\columnwidth, keepaspectratio]{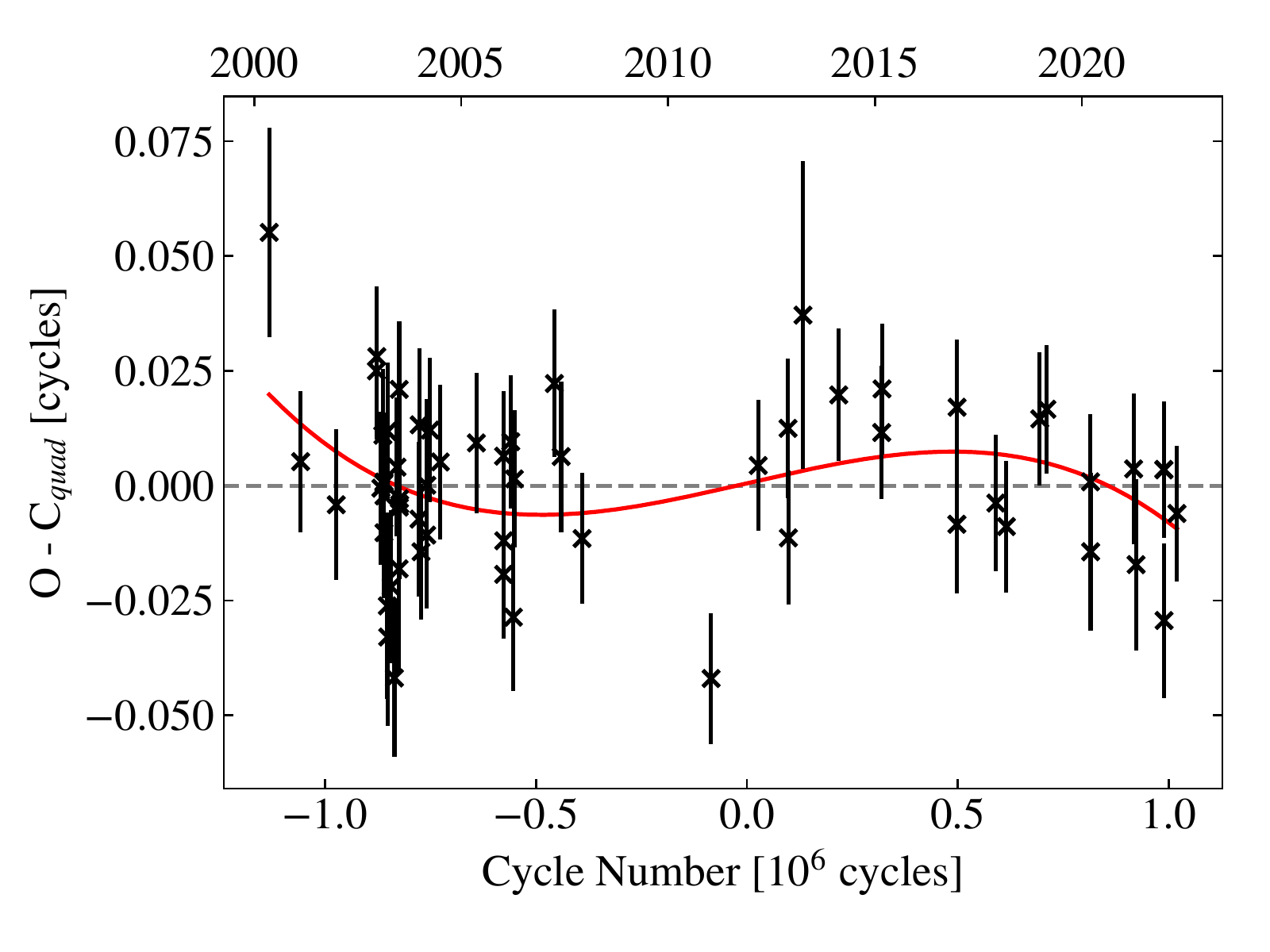}
    \caption{\emph{Left:} Timing residuals, observed minus calculated, relative to a linear ephemeris $\phi_0 = 11.1$ cycles, $f_0=3.11021829$\,mHz, T$_0=55849$ BJD~TDB. \emph{Right:} Timing residuals relative to the quadratic ephemeris given in Table~\ref{tab:ephemerides}. The solid curved line in the right-hand panel shows a cubic ephemeris fit offset relative to the same quadratic ephemeris. Error bars are included in both plots, but are too small to be seen in the linear ephemeris panel. The residuals are expressed in terms of orbital cycles. The curvature in the linear ephemeris panel is such that HM~Cnc has now advanced by more than 80 entire orbits compared to the period it had at the time of its discovery. The top axes are labelled by the Gregorian year of observation. Multi-filter timing solutions on individual nights are combined to represent a single data point, including a $0.014$~cycles error added in quadrature to account for system variability (\cref{subsec:Flickering}). This was set so that the $\chi^2$ per degree of freedom of the cubic fit equals one. A tabulated list of individual-filter timing solutions is available in Appendix~\ref{tableappendix:timingSolutionsAll}. Horizontally drawn in grey on the left-hand panel are the spans of cycles covered by X-ray observations. These overlap with our observations and are discussed in \cref{subsec:Flickering}.}
    \label{fig:O-C}
\end{figure*}
\begin{table*}
\centering
\caption{Our ephemerides for a cubic and quadratic ephemeris, solved with equation~(\ref{eqn:PhasingSol}). All errors are quoted to a 1$\sigma$ uncertainty. The epoch used in these fits was 55849, BJD~TDB. This solution is representative of our $u'-g'$ and $r'-g'$ phase offset correction (Table \ref{tab:offsets}) and $0.014$~cycle flickering correction to the error of each night in quadrature (\cref{subsec:Flickering}).}
\label{tab:HMCnc_solution}
\centering
\begin{tabular}{l c c}
\hline
Parameter & Cubic & Quadratic \\
\hline
Phase Offset, $\phi_0$ (Cycles) & $0.611\pm0.004$ & $0.611\pm0.004$\\
Orbital Frequency, $f_0$ (Hz) &  $0.003110224895\pm30\,$pHz  & $0.003110224829\pm 10\,$pHz \\
Frequency Derivative, $\dot f_0$ (Hz$\,$s$^{-1}$)& ($3.538\pm0.001$) $\times 10^{-16}$  & ($3.538\pm0.001$) $\times 10^{-16}$ \\
Frequency Second Derivative, $\ddot f_0$ (Hz$\,$s$^{-2}$) & ($-5.38\pm2.10$) $\times 10^{-27}$ & --\\
\hline
\label{tab:ephemerides}
\end{tabular}
\end{table*}

We fit Fourier series solutions to the extracted photometry of the form
\begin{equation}
d + A \cos\left(\omega (t-t_0)\right) + B \sin\left(2\omega (t-t_0)\right) + C \cos\left(2\omega (t-t_0)\right),
\label{eqn:WaveFit}
\end{equation}
where $\omega = 2\pi f$ is the orbital angular frequency, $t$ is the mid-exposure time of an individual observation, $t_0$ is the timing solution and is the time that we seek to measure, $d$ is a normalised flux offset, and $A$, $B$ and $C$ are amplitudes of the fundamental and first harmonic; the harmonic terms account for the non-sinusoidal variability of HM~Cnc in the optical (as noticeable in Fig.~\ref{fig:LCs}). For each night of data with each filter, $d$, $A$, $B$, $C$ and $t_0$ were allowed to vary while $\omega$ was held fixed at the expected angular frequency at the time of observation from the cubic ephemeris of \citet[][]{Strohmayer2021HMCnc}. Including further harmonics to equation~(\ref{eqn:WaveFit}) generated a negligible change in the timing measurement. Uncertainties on $t_0$ were deduced by taking the standard deviation of 1000 bootstrapping iterations, as used by \citet{Pelisoli2021FastSpinningWD}. Example Fourier series solutions and discussion on specific cases can be later found in \cref{subsec:Flickering}.

For identical filters, if a periodic signal was difficult to detect (e.g. due to poor seeing, short observing time on target), we used the Fourier series solution of the night with the longest observation time for a given filter to search for a timing solution again. We note that there was little difference in $t_0$ if we allowed the parameters of $A, B$ and $C$ to be free or if we fixed them using a solution to the combined filter dataset. This indicates that there has been no/little detectable change in the periodic signal over the full duration of observations, which may have been induced by a changing accretion spot location. We also note that there was no obvious change in the shape of the Fourier series solution about the peak X-ray flux when comparing nights of observation for identical filters spaced years apart.

We then searched for any filter-dependent offsets to the final timing measurements and present these computed offsets in Table \ref{tab:offsets}. Only the $u'-g'$ and $r'-g'$ offsets indicate a relevant deviation from a zero-phase offset at the 1$\sigma$ level, and so are corrected for all ULTRACAM data using the $u'$ and $r'$ filters. All other observations from data that we have obtained or from archival sources are considered to have a phase offset of zero.
\begin{table}
    \caption{Phase offsets for each filter used in this study for simultaneous multi-filter observations. The $g_s$ or $g'$ bands are used as a reference.}
    \centering
    \begin{tabular}{c c c c}
        \hline
        Filter & $\Delta\phi$ (cycles) & Filter & $\Delta\phi$ (cycles) \\
        \hline
         u$_s$ $-$ g$_s$ & ~0.0015 $\pm$ 0.0040 & u' $-$ g' & ~0.0087 $\pm$ 0.0022\\
         r$_s$ $-$ g$_s$ & -0.0006 $\pm$ 0.0029 & r' $-$ g' & -0.0023 $\pm$ 0.0016  \\
         i$_s$ $-$ g$_s$ & ~0.0019 $\pm$ 0.0047 & i' $-$ g' & ~0.0001 $\pm$ 0.0090\\
         z$_s$ $-$ g$_s$ & ~0.0020 $\pm$ 0.0128\\
         \hline
    \end{tabular}
    \label{tab:offsets}
\end{table}

\subsection{Constraining the ephemeris}
With the full set of filter-offset-corrected (and flickering-considered, see~\cref{subsec:Flickering}) timing solutions, we proceeded by assigning each an integer cycle number, as inferred using the ephemerides of \citet{Barros2007HMCnc} and \citet{Strohmayer2021HMCnc} while correcting for our own epoch, $T_0$. We chose an epoch $T_0=55849$, BJD~TDB, as we found that this gave the smallest correlation between the fitting parameters and is located near the centre of the full observing span as expected. Cycle numbers from both of these ephemerides gave the same integer cycle number for all nights and inferred cycle numbers were manually verified for observations taken more recently than those in \citet[][]{Strohmayer2021HMCnc}. In no case was a jump in cycle number observed, which would be easily noticeable since typical uncertainties are much less than a cycle in magnitude.

It is clear (see Fig.~\ref{fig:O-C}) that HM~Cnc deviates by multiple cycles with respect to a linear fit that has a constant orbital frequency, so we added higher order coefficients to compute a quadratic and then a cubic fit. The set of cycle numbers were fit using
\begin{equation}
    \phi\left(t\right) = \phi_0 + f_0 \left(t-T_0\right) + \frac{\dot f_0}{2}\left(t-T_0\right)^2 + \frac{\ddot f_0}{6}\left(t-T_0\right)^3
\label{eqn:PhasingSol}
\end{equation}
with $\phi(t)$ the cycle number, $\phi_0$ a cycle offset at the epoch $T_0$, $f_0$ the frequency at the epoch $T_0$, $\dot f_0$ its first derivative and $\ddot f_0$ its second. For a linear fit, the first two terms on the right-hand side of equation~(\ref{eqn:PhasingSol}) apply, the first three terms for a quadratic and all four for a cubic. The quadratic term corresponds to the parabola that spans 21 cycles from minimum to maximum (Fig.~\ref{fig:O-C}) which means that HM~Cnc has advanced by over 80 orbital cycles relative to the period that it had upon discovery, the result of gravitational-wave driven inspiral.

A fit to our timing measurements was performed using a least-squares analysis through the \textsc{scipy lstsq} solver in the \textsc{scipy} linear algebra package, which we also used to return $1 \sigma$ statistical errors on all fitted parameters. In the process, the errors on observations were inflated in quadrature so that the reduced $\chi^2$ is equal to one, where flickering in the system largely encourages the inclusion of this procedure (as discussed in \cref{subsec:Flickering}). Our residuals for solutions with a linear and quadratic ephemeris are presented in Fig.~\ref{fig:O-C} with the ephemerides for a quadratic and cubic model given in Table~\ref{tab:HMCnc_solution}. Corrections of the ephemerides due to the Shklovskii effect \citep[][]{Shklovskii1970effect} and galactic rotation are negligible (see Appendix~\ref{secappendix:ephemerisCorr}). The $\dot f_0$ coefficient of our quadratic ephemeris is precise to a 0.03\% level and is 40 times more precise than that determined by \citet[][]{Barros2007HMCnc}, also derived from optical data.

To check if the $\ddot f_0$ term is statistically significant, we performed an $F$-test between the quadratic and cubic fits. Under the hypothesis that the cubic term is not a significant contribution to the ephemeris, the following $F$-ratio for $n$ data points
\begin{equation}
    F_{\textrm{ratio}} = \left(\chi^2_{\textrm{quad}} - \chi^2_{\textrm{cubic}}\right)  \left(\frac{\chi^2_{\textrm{cubic}}}{n-4}\right)^{-1}
\end{equation}
\noindent is expected to have an $F$(1, $n-4$) distribution, where the 1 in the parenthesis represents the difference in the number of degrees of freedom between the quadratic and cubic fits and $n-4$ reflects the unique degrees of freedom in the cubic fit. We compute the $F$-ratio to be 6.08 for our $n=60$ measurements. The cumulative distribution function of an $F$-distribution can be used to infer at what significance the null-hypothesis that a cubic term is not reflective of the data can be rejected. As such, the cubic term (and so $\ddot f_0$) is significant at the 98.3\% level (which is significant at 95\%, or 2$\sigma$, but not at 99.7\%, or 3$\sigma$). We also carried out an $F$-test on the digitised X-ray data between the quadratic and cubic solutions specified in \citet[][]{Strohmayer2021HMCnc} to test the significance of $\ddot f$ (see \cref{subsec:Flickering}). The reduced $\chi^2$ of a cubic fit to the X-ray data was 1.04, so we again inflated the error of all data points in quadrature to gain a reduced $\chi^2$ equal to one. We found $\ddot f = -9.41\pm1.42\times 10^{-27}\,$Hz$\,$s$^{-2}$ and an $F$-ratio of 44.1 for the X-ray data. From the cumulative distribution of $F$(1,172), this represents a near 100\% significance. Though with sparser sampling, individual measurements for the X-ray data have higher precision since the peak flux per cycle is better defined by a sharp rise towards an intensity maximum, while the optical is more susceptible to variability. Taking the $F$-test results as evidence that $\ddot f$ is a true characteristic of the system, all further mentions of $f_0$, $\dot f_0$ and $\ddot f_0$ in this paper assume the cubic solution of Table~\ref{tab:ephemerides}.

Our optical measurements of $\dot f_0$ and $\ddot f_0$ allow us to estimate the maximum frequency of HM~Cnc before a flip of sign of $\dot f$ is observed. If $\ddot f_0$ remains constant, the system frequency would be expected to evolve as
\begin{equation}
    f(t) = f_0 + \dot f_0 (t-T_0) + \frac{\ddot f_0}{2} (t-T_0)^2
    \label{eqn:FrequencyEvolution}
\end{equation}
\noindent where a maximum frequency is reached when $t=-\dot{f_0}/\ddot{f_0}+T_0$. We would thus predict the frequency maximum to occur in approximately $2100\pm800$\,yrs from now. The timing of this event is consistent with the frequency maximum predicted in \citet[][]{Strohmayer2021HMCnc}. As always with timing studies, a larger baseline of data would improve the precision of the ephemeris and, in particular, better constrain $\ddot f_0$.

Interestingly, our $\ddot f_0$ is of opposite sign to that predicted in earlier literature, with \citet{Deloye2006predictingFddotFromLacc} expecting a \textit{positive} $\ddot f$ of magnitude 10$^{-28}\,$Hz$\,$s$^{-2}$. It would be expected that $\ddot f$ is positive for almost the entirety of inspiral until merger under purely general relativistic orbital decay, unless a re-stabilisation of the orbit due to mass transfer occurs. Consequently, it is thought that HM~Cnc is in a mass transfer turn-on phase that will drive the system to smaller orbital frequencies (longer periods) following the frequency maximum.

Under solely gravitational-wave orbital angular momentum loss
\begin{equation}
    \centering
    \dot f_{GW} = \frac{96}{5}~\pi^{8/3}~\left( \frac{G\mathcal{M}}{c^3}\right)^{5/3}~f_{GW}^{11/3}
    \label{eqn:fdotGR}
\end{equation}
\noindent with $f_{GW}$ the frequency of emitted gravitational waves (equal to twice the orbital frequency), $G$ the gravitational constant, $c$ the speed of light and $\mathcal{M}~=~(m_a\,m_d)^{3/5}~/~(m_a+m_d)^{1/5}$ the chirp mass. For our measured $f_0$ and $\dot f_0$, we obtain an observed chirp mass of $0.3203\pm0.0001\,$\(\textup{M}_\odot\). Given that the trajectory of HM~Cnc is impacted by the mass accretion rate (which acts to oppose inspiral) and not from gravitational wave radiation alone, this observed chirp mass is a lower bound for the true chirp mass of the system.

\subsection{Flickering}
\label{subsec:Flickering}
\begin{figure}
    \centering
    \includegraphics[trim={0.5cm 0.5cm 0.5cm 0.cm},clip,width=\columnwidth, keepaspectratio]{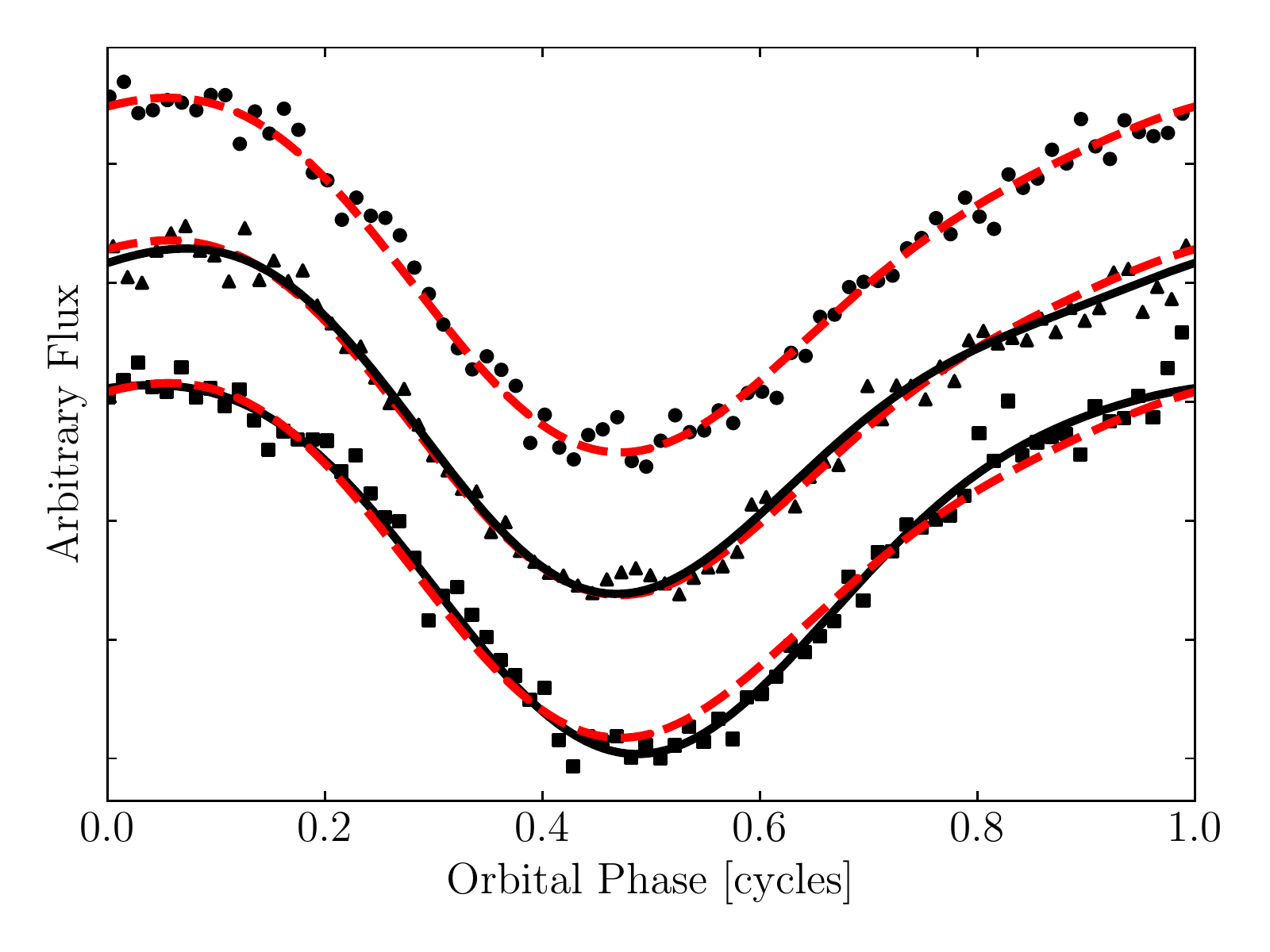}
    \caption{Solutions to VLT g-band data, with each plotted light curve representing (binned) data of the first hour of VLT observations each night. Bottom: 2005-11-26 (squares), middle: 2005-11-27 (triangles), Top: 2005-11-28 (circles). The normalised flux is offset between individual nights for clarity. The time at phase 0.0 represents the extracted $t_0$ for each hour of observations. The Fourier solution for the night of 2005-11-28 is shown in dashed red. The Fourier solutions for other nights are displayed in solid black.}
    \label{fig:FlickeringGraph}
\end{figure}

We noticed variations in the timing solutions of HM~Cnc between adjacent nights with ideal conditions. For data from the VLT with ULTRACAM, observations were performed for a similar duration at similar airmasses over the three nights observed. The observations had a seeing~<~1.0\arcsec and multiple consecutive hours spent on target, resulting in high signal-to-noise ratio (SNR) data with no clear outliers. Fig.~\ref{fig:FlickeringGraph} depicts the changes in the Fourier series solution for sets of observations that were cut to have a one hour duration, with phase-folded light curves included. There are noticeable differences between the solutions of different nights, particularly evident when comparing the phase of minimum flux. Analysis of the full set of VLT nights over an hourly timescale also indicated that there are small changes between solutions with the longest single observing period being three hours, though nightly differences were more apparent. A similar impact to the Fourier solutions and the returned timing solutions was apparent in the (simultaneously observed) u'- and r'-band VLT observations. When we compared the best-quality adjacent nights of NOT data (BJD=52645--52647) each with over 3 hours on target per night, we saw hourly light curve fluctuations at a similar level, though with larger error in individual measurements due to the smaller telescope aperture. We presume that a changing apparent light-curve morphology is a consequence of changes in the system mass transfer rate, creating a flickering effect typical of accreting binaries, while the long-term light-curve morphology is less impacted.

A changing optical periodic signal inherently limits the accuracy of all timing solutions, where, taking the mean of the VLT/ULTRACAM nights to be a reference point, the scatter in $t_0$ between adjacent nights could be as much as $\pm$0.01--0.015~cycles for all bands. In this study, we thus combine timing solutions from all filters (phase offset corrected where applicable) to obtain a single timing measurement for the night of observations. Then, we add a fixed error in quadrature due to flickering to each night's timing measurement, such that the reduced $\chi^2$ of our ephemerides is equal to one. Overall, this error adjustment was 0.014~cycles.

Also in support of system flickering, Fig.~7 of \citet[][]{Strohmayer2021HMCnc} compares phase-folded X-ray flux profiles from the Chandra and NICER datasets. A clear difference in flux profile is noticeable after the X-ray peak flux as the flux is decreasing, and to a lesser extent at the peak X-ray flux. This shows deviation over a multiple year time-span, but it is possible that these changes occur on much shorter timescales, e.g. hours.

\begin{figure}
    \centering
    \includegraphics[trim={0.cm 0.cm 0.cm 0.05cm},clip, width=\columnwidth, height=6.59cm]{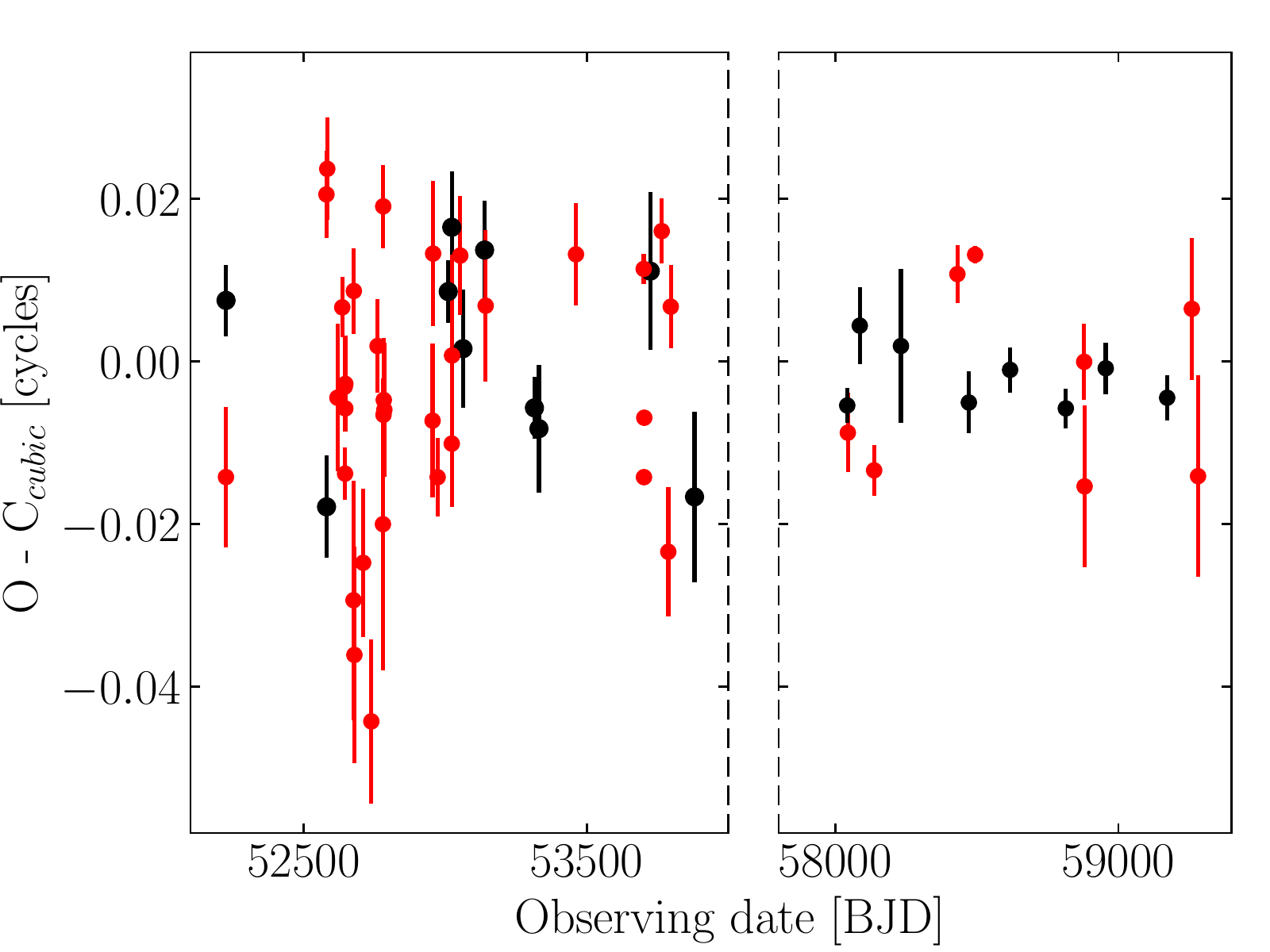}
    \caption{A comparison of the X-ray timing solutions (black) of \citet[][]{Strohmayer2021HMCnc} with our optical measurements (red). \citet[][]{Strohmayer2021HMCnc} measurements have been binned to represent a single measurement for one observing session, lasting hours at each time. X-ray data on the left of the divide are from Chandra, and on the right are from NICER, with broken axes represented by the dashed line between the plots. The scale is the same in both. Optical measurements within the broken axes are not included in the figure, though are plotted in Fig.~\ref{fig:O-C} and comprise a further eight years of coverage. Raw errors are included without the `flickering correction' described in \cref{subsec:Flickering}. O-C$_{\textrm{cubic}}$ was calculated using our cubic ephemeris in Table~\ref{tab:HMCnc_solution} for both the optical and the X-ray measurements.}
    \label{fig:XrayVsOptical}
\end{figure}

We then set out to determine whether the flickering was apparent in the X-ray timing solutions as well by comparing our data to the nearest X-ray observation. To do this, we utilised the data from Fig.~6 of \citet[][]{Strohmayer2021HMCnc} by digitising and extracting the plotted data points, followed by a conversion of each to an observed time of peak X-ray flux. After, we compared our own timing solutions with the X-ray times for Chandra and NICER data separately. A phase difference between the time of X-ray peak flux and our derived solutions of $t$ was compensated for by minimising the $\chi^2$ between our optical timing solutions and the X-ray for our cubic-fit ephemeris given in Table~\ref{tab:ephemerides}, generating a relative phase difference $\delta\phi\,\approx\,0.14$ cycles from the optical to X-ray. A comparison of the two sets of solutions is plotted in Fig.~\ref{fig:XrayVsOptical}, where the 176 individual X-ray measurements have been combined if they coincide with half of a day of each other.

Very few optical/X-ray observations were taken within a day of each other, making a direct comparison difficult. Little scatter is seen in both datasets from the most recent observations, while the earliest show a similar scatter except for the first optical/X-ray measurements. All that we can assert from the data is that an impact on the accuracy of timing solutions at any wavelength cannot be ruled out.


\section{Evolutionary Modelling}
\label{sec:EvolutionaryModellingMESA}
\begin{figure}
    \centering
    \includegraphics[trim={0.5cm 0.5cm 0.5cm 0.cm},clip,width=\columnwidth, keepaspectratio]{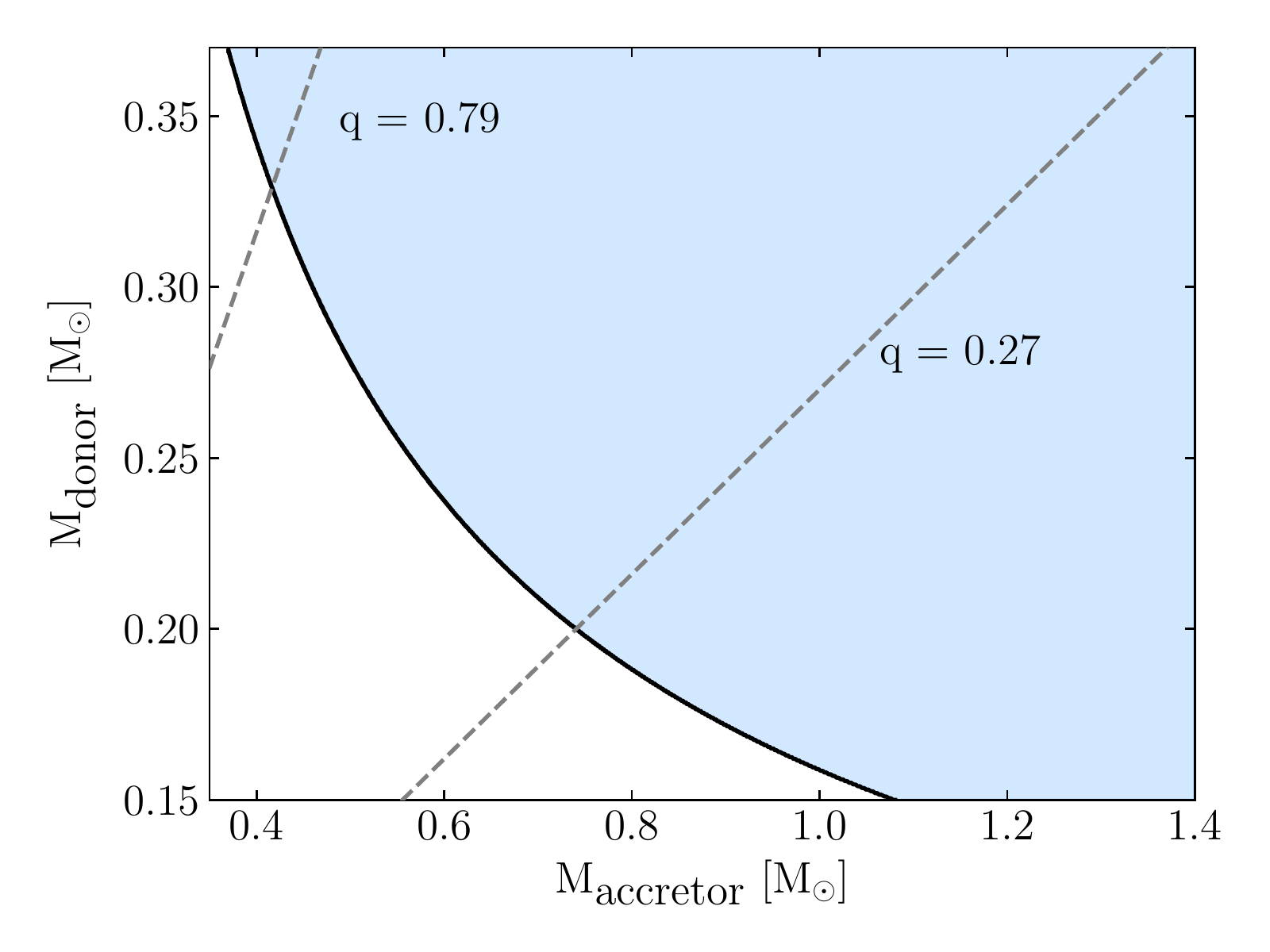}
    \caption{Permitted mass combinations for the stars of HM~Cnc, given by the shaded blue region. The black solid line represents valid solutions for our observed chirp mass assuming that the orbital decay is solely due to general relativistic effects. Mass transfer permits a smaller $\dot f_0$ than purely general relativistic orbital decay in a detached system, meaning that larger system chirp masses (and thus donor and accretor masses) than indicated by the observed chirp mass are possible. Reported boundaries of the mass ratio, $q=M_{\textrm{donor}}/M_{\textrm{accretor}}$ from \citet{Roelofs2010HMCncMassRatio} are drawn as dashed grey lines.}
    \label{fig:Allowed_masses}
\end{figure}

Our measured $\ddot f_0$ combined with the $\dot f_0$ and $f_0$ is a very strong constraint upon HM~Cnc's evolution. We explored simulated binary evolution models with comparison to our observed ephemeris using the Modules for Experiments in Stellar Astrophysics (MESA) 1-D stellar evolution code \citep{Mesa1, Mesa2, Mesa3binary, Mesa4, Mesa5}, release version 15140, to investigate the AM~CVn DWD configuration. It has been proposed by \citet{DAntona2006} that the donor of HM~Cnc may be an extremely low mass (ELM) WD that is not fully degenerate and has a thick hydrogen layer, which allows the period to shorten even though accretion occurs. The Lyman-$\alpha$ detection in our HST spectra supports this theory, as well as larger mass, fully-degenerate WD donors with hydrogen in their envelope. Both types of situations are included in our MESA models. Plotted in Fig.~\ref{fig:Allowed_masses} are the valid mass combinations for each star that our observed chirp mass permits, which we surveyed in the MESA models.
\subsection{Building a Donor}
\label{subsec:buildingDonor}
We approximate the full binary evolution by constructing a He WD from an adaptation of the \textsc{\text{make\_he\_wd}} MESA test suite. The way this is performed is by evolving a pre-main sequence to a zero age main sequence star, and then by allowing fusion to take place until the core mass reaches the desired mass of a He WD minus the mass of the envelope. We then cut the evolution and gently remove the envelope of the star until only the He core resides. Next, we evolve the star to an effective temperature of 10\,kK and then gradually accrete pure hydrogen onto the surface, giving us full control of the total hydrogen envelope mass. What is left is a He core WD with a thin, pure hydrogen layer, which we evolve for 0.5$\,$Gyrs while enabling diffusion, stratifying the layers of the WD to leave a realistic structure\footnote{Our WD models have Z=0.00 due to slow convergence on a solution with our crafted structure in the diffusion process. A couple of test cases of Z=0.02 were evolved as an AM~CVn to ensure fair comparisons, where the impact of metallicity in the accreted material was found to be negligible.}.

The maximum hydrogen envelope mass in our models was inferred from the Z=0.0001 He WD models of \citet[][]{Istrate2016}, choosing models where no hydrogen flash has occurred such that the star maintains a thick hydrogen envelope\footnote{Though \citet[][]{DAntona2006} suggest that larger hydrogen envelope masses of 10$^{-2}\,$\(\textup{M}_\odot\) could be required for the model, we consider the non-hydrogen-flash \citet[][]{Istrate2016} models to have an improved maximum hydrogen envelope mass.}. For smaller or larger mass WDs than given in the range of models provided by \citet[][]{Istrate2016}, we linearly extrapolated their models, with the largest hydrogen envelope we used being M$_\textrm{H}=6.5\times10^{-3}\,$\(\textup{M}_\odot\) for models of a 0.16$\,$\(\textup{M}_\odot\) donor.

While the full set of \citet[][]{Istrate2016} models serves as a good approximation of the amount of hydrogen in the He WD's envelope, the evolution of a compact DWD is very different to their modelling of a neutron star companion to a WD. This, as well as potential hydrogen flashes, suggests that the initial hydrogen envelope could be less than these models. With that in mind, we investigated the compatibility of different masses of hydrogen envelopes by considering our own set of He WD models, whereby a minimum donor hydrogen envelope consists of M$_\textrm{H}=0.1\times10^{-3}\,$\(\textup{M}_\odot\). We sampled hydrogen envelope masses between these limits also, computing models for M$_\textrm{H}=1.0\times10^{-3}\,$\(\textup{M}_\odot\) and M$_\textrm{H}=3.0\times10^{-3}\,$\(\textup{M}_\odot\).

\subsection{Binary Modelling}
\label{subsec:BinaryModelling}
We modelled HM~Cnc by starting the binary evolution with the fabricated donor He~WD and a point mass accretor before any Roche lobe overflow has begun (typically with an initial period of 0.05 days). It is worth noting that \citet[][]{Wong2021AMCVnMESAPureHe} have modelled AM~CVn systems with MESA by assuming a pure He donor before any mass transfer and fully evolve two stars rather than one star and a point mass. This would be advantageous here also, however the fact that hydrogen is present in the spectrum and that the hydrogen fraction largely controls the timing of maximum orbital frequency encourages us to include the hydrogen envelope. Though evolving two stars would still be possible, computations including hydrogen to incorporate nuclear burning on the accretor's surface are extremely computationally expensive. Our need to precisely control initial conditions, to simulate a wide grid of models and the fact that nuclear burning on the accretor's surface is important at the onset of mass accretion but less so later in the evolution \citep[][]{Kaplan2012OrbitalEvolutionOfCompactWDBinaries} led us to proceed with the point mass accretor abstraction, as was similarly initiated by \citet[][]{Chen2022MESAamcvnWDdonor}. The abstraction ignores the possible impact of thermonuclear runaways on the surface of the WD which could result in enhanced angular momentum and mass loss \citep[][]{Shen2015EveryDWDmerge}. Given that we know that HM~Cnc has not merged and that most of the donor hydrogen envelope has been depleted, the binary has survived hydrogen-rich nova episode(s) and our models are only susceptible to complications from a helium nova episode \citep[][]{Shen2015EveryDWDmerge}, which may occur in the future. This makes the evolution of our models realistic leading up to $f_0$, $\dot f_0$ and $\ddot f_0$, even though the events of the future are uncertain.

Most importantly to the situation at hand, our dynamics revolve around the framework of the Roche lobe approximation of \citet[][]{Eggleton1983} and mass transfer rates following the method outlined in \citet{Ritter1988}. We evolve the orbital angular momentum, $J_{\textrm{orb}}$, according to
\begin{equation}
    \dot J_{\textrm{orb}} = \dot J_{\textrm{gr}} + \dot J_{\textrm{ml}} + \dot J_{\textrm{mb}} + \dot J_{\textrm{ls}} 
    \label{eqn:Jorb}
\end{equation}
where computed values for the rate of orbital momentum change due to gravitational wave radiation, $\dot J_{\textrm{gr}}$, stellar wind mass loss, $\dot J_{\textrm{ml}}$, magnetic breaking, $\dot J_{\textrm{mb}}$, and spin-orbit (LS) coupling, $\dot J_{\textrm{ls}}$, are handled by MESA according to the methods outlined in \citet{Mesa3binary}. Mass transfer between the two stars is considered to be perfectly efficient and we apply a tidal synchronisation following the prescription of \citet[][]{Hurley2002BinaryEvolution} using inbuilt MESA routines.

When an accretion stream feeds into a disc around the accretor, all transferred orbital angular momentum is assumed to return to the orbit of the system; there is no contribution to equation~(\ref{eqn:Jorb}). When accretion occurs via direct impact, this condition no longer holds. To account for this, we added an extra angular momentum sink to the MESA calculation, following the prescription of \citep[see e.g.][]{Marsh2004WDMassTransfer}
\begin{equation}
    \dot J_{\textrm{di}} = J_{\textrm{orb}} \sqrt{r_h~(1+q)} \frac{\dot M_d}{M_d}
    \label{eq:Jdi}
\end{equation}
\noindent where $\dot J_{\textrm{di}}$ is the extra loss of orbital angular momentum due to direct-impact mass transfer, $r_h$ follows the prescription of equation~(13) of \citet{Verbunt1988rh} as the equivalent radius of the orbiting accreted material and $q=M_{\textrm{d}}/M_{\textrm{a}}$ is the mass ratio. We determined the radius of the star at which an accretion stream would pass by, resulting in disc-fed accretion, using equation~(6) of \citet[][]{NelemansII2001populationSynthesisOfWDsAMCVn}. Given that we consider a point mass accretor, its radius was computed using a mass-temperature-radius relationship (MTRR) from the Montreal CO~WD models \citep[][]{Bedard2020MontrealWDModelsMTR}, where we considered the temperature of the accretor to be 42\,500K from our spectrum fit (\cref{subsec:spectroscopy}). Hence, we have direct impact accretion if the MTRR radius of the accretor is larger than the radius required for the accretion stream to pass by, or disc-fed accretion on the contrary. In our simulations, this accretor radius is a minimum since the star is likely inflated from mass transfer and partially from irradiation, such that direct impact could begin slightly earlier. For each timestep and model simulated in MESA, we computed whether matter is accreted via direct impact or not and accordingly turned on/off the inclusion of equation~(\ref{eq:Jdi}) to the computed $\dot J_{\textrm{orb}}$ in equation~(\ref{eqn:Jorb}).

Diffusion processes were continuously modelled while evolving the binary, before and during mass transfer. The time-step integration itself was variable in the models; longest initially when there is no mass accretion and smallest when the two stars are at their closest separation. The mass transferred in each time-step was calculated implicitly. A test for a merger was analysed for every model with every time-step integration, however unstable mass transfer only occurred for the less-massive-accretor models in the range permitted in Fig.~\ref{fig:Allowed_masses}.
\begin{figure*}
    \centering
    \includegraphics[width=\columnwidth, keepaspectratio]{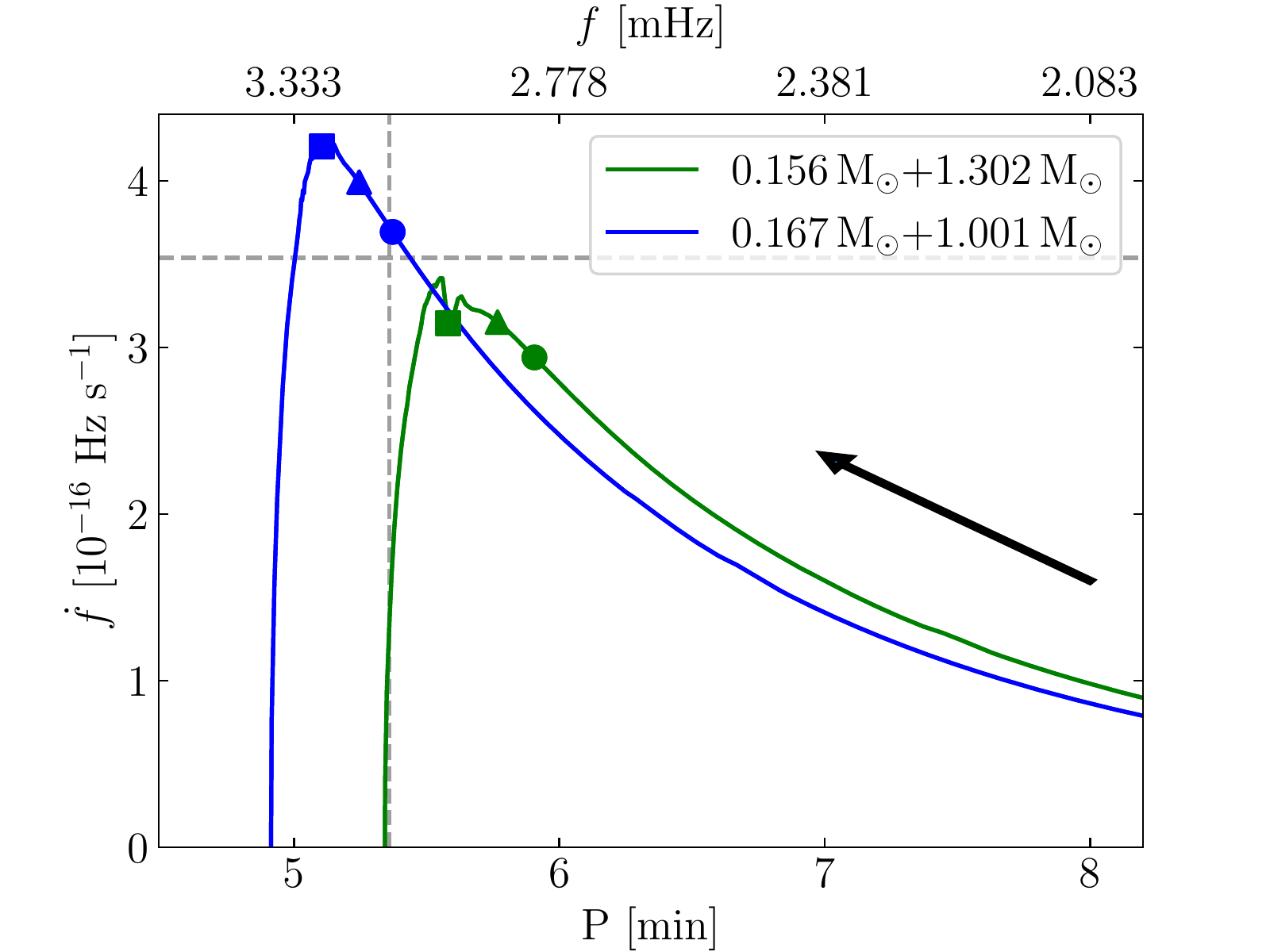}
    \includegraphics[width=\columnwidth, keepaspectratio]{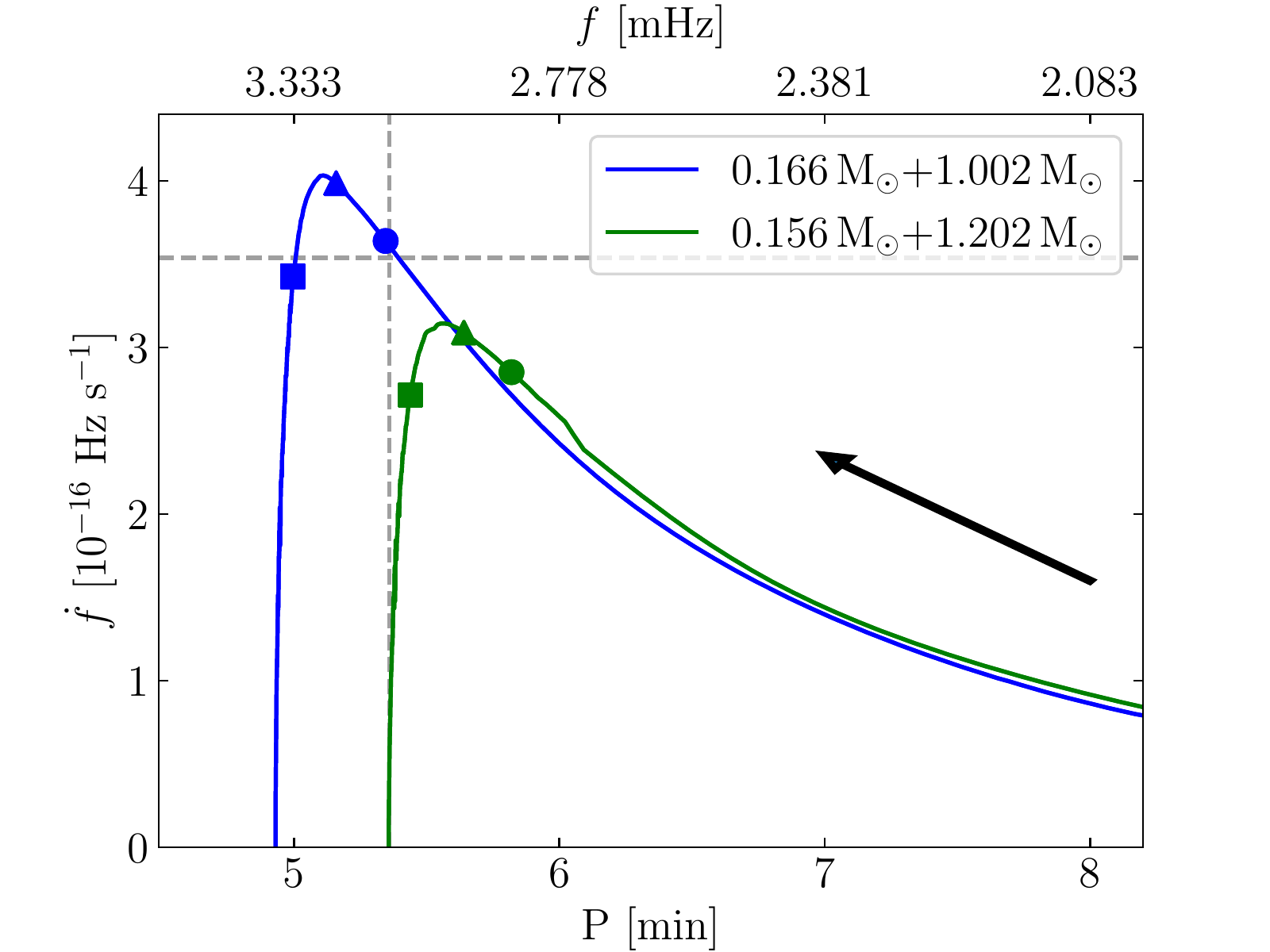}
    \includegraphics[width=\columnwidth, keepaspectratio]{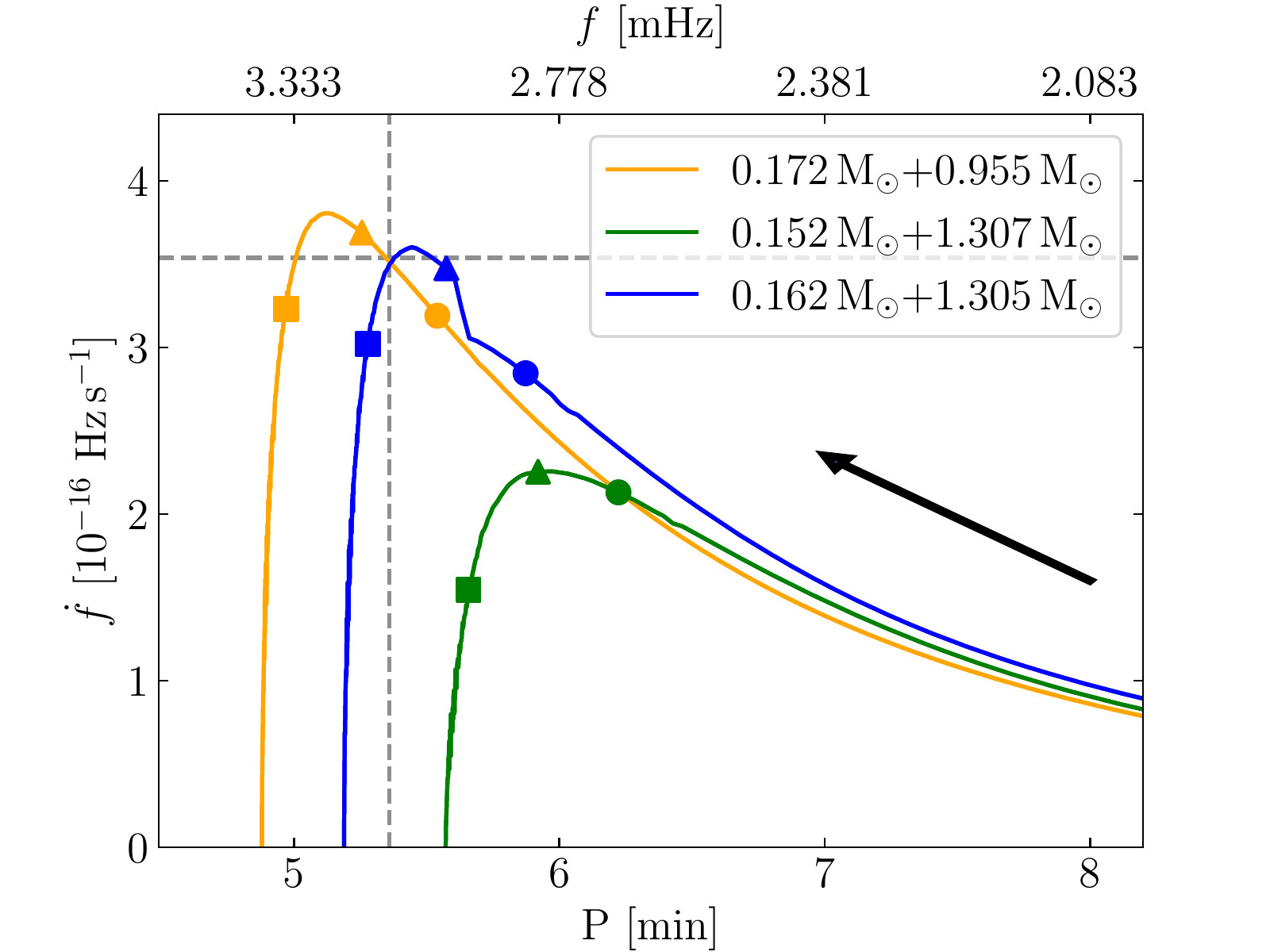}
    \includegraphics[width=\columnwidth, keepaspectratio]{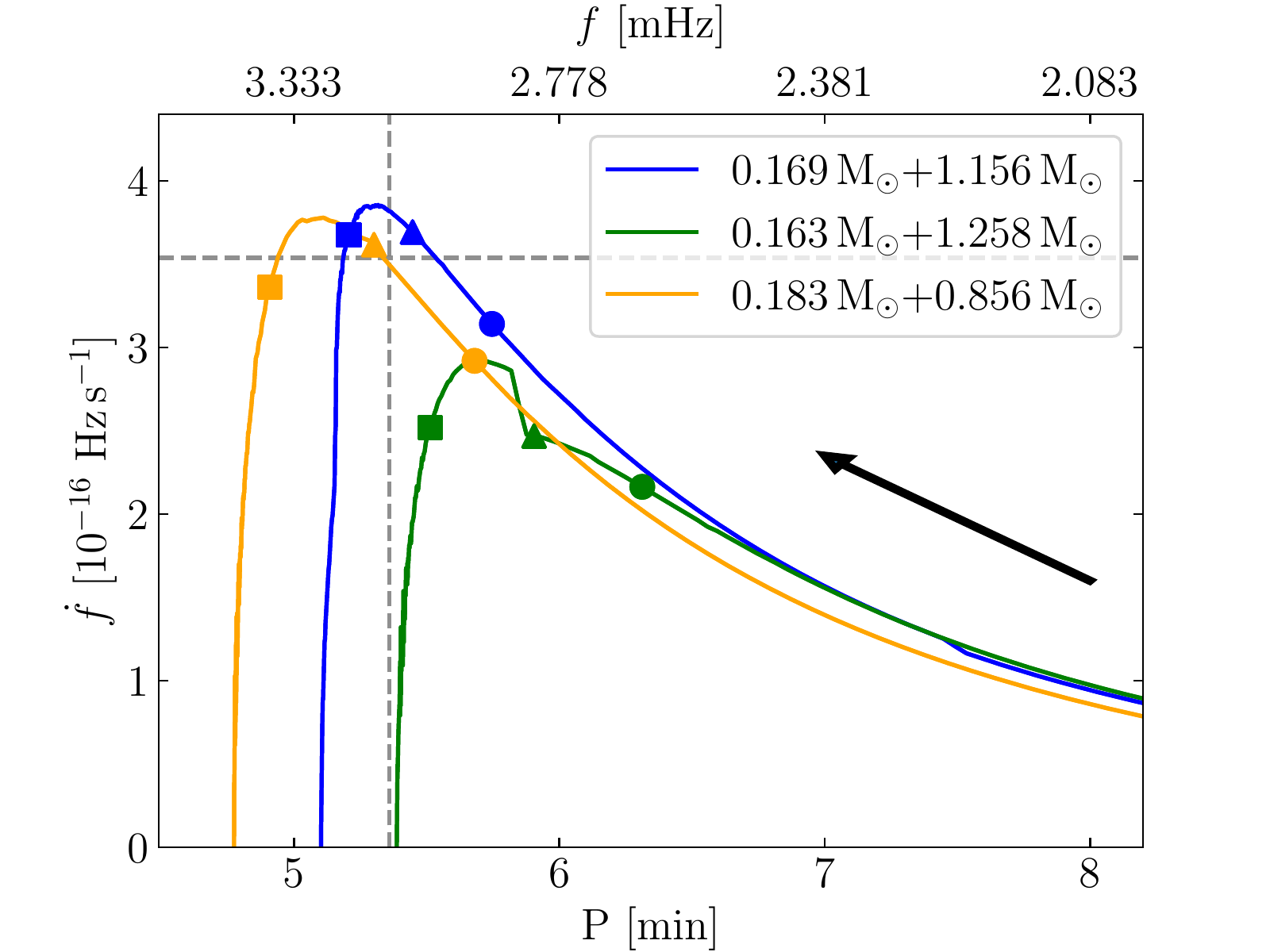}
    \caption{The best-fitting MESA results for HM~Cnc. Top left M$_\textrm{H}~=~0.1\times10^{-3}\,$\(\textup{M}_\odot\); top right M$_\textrm{H}~=~1.0\times10^{-3}\,$\(\textup{M}_\odot\); bottom left M$_\textrm{H}~=~3.0\times10^{-3}\,$\(\textup{M}_\odot\); bottom right M$_\textrm{H}$ of \citet[][]{Istrate2016} models (see text in \cref{subsec:buildingDonor}). Black arrows show the direction of evolution; increasing in time. Coloured circles, triangles and squares represent the moments at which the hydrogen envelope fraction of the donor, H/He (by number), equals 0.1, 0.05 and 0.01 respectively. The dashed grey crosshairs represent the values of our observed $f_0$ and $\dot f_0$. The legend of each plot displays the predicted model masses of the donor and the accretor when the binary has a negative $\ddot f_0$. `Kinks' in the direction of evolution occur due to a switch from disc to direct-impact accretion in certain cases. All models have a donor temperature in the range of 5.5--6.0\,kK at present day. In each plot, the most massive donor models reach the largest maximum frequencies before turn-around.}
    \label{fig:MESAresults}
\end{figure*}
\begin{figure*}
    \centering
    \includegraphics[width=\columnwidth, keepaspectratio]{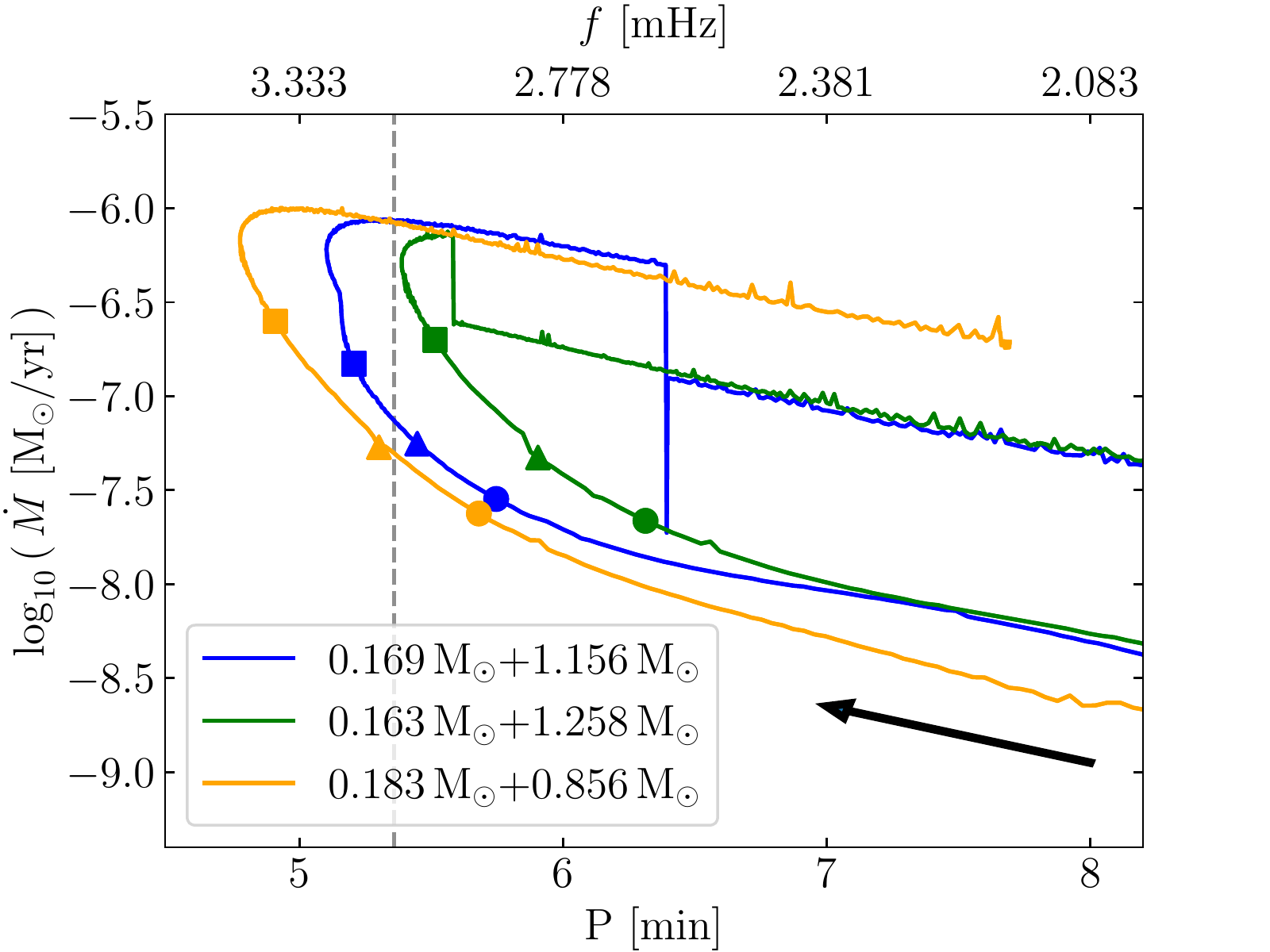}
    \includegraphics[width=\columnwidth, keepaspectratio]{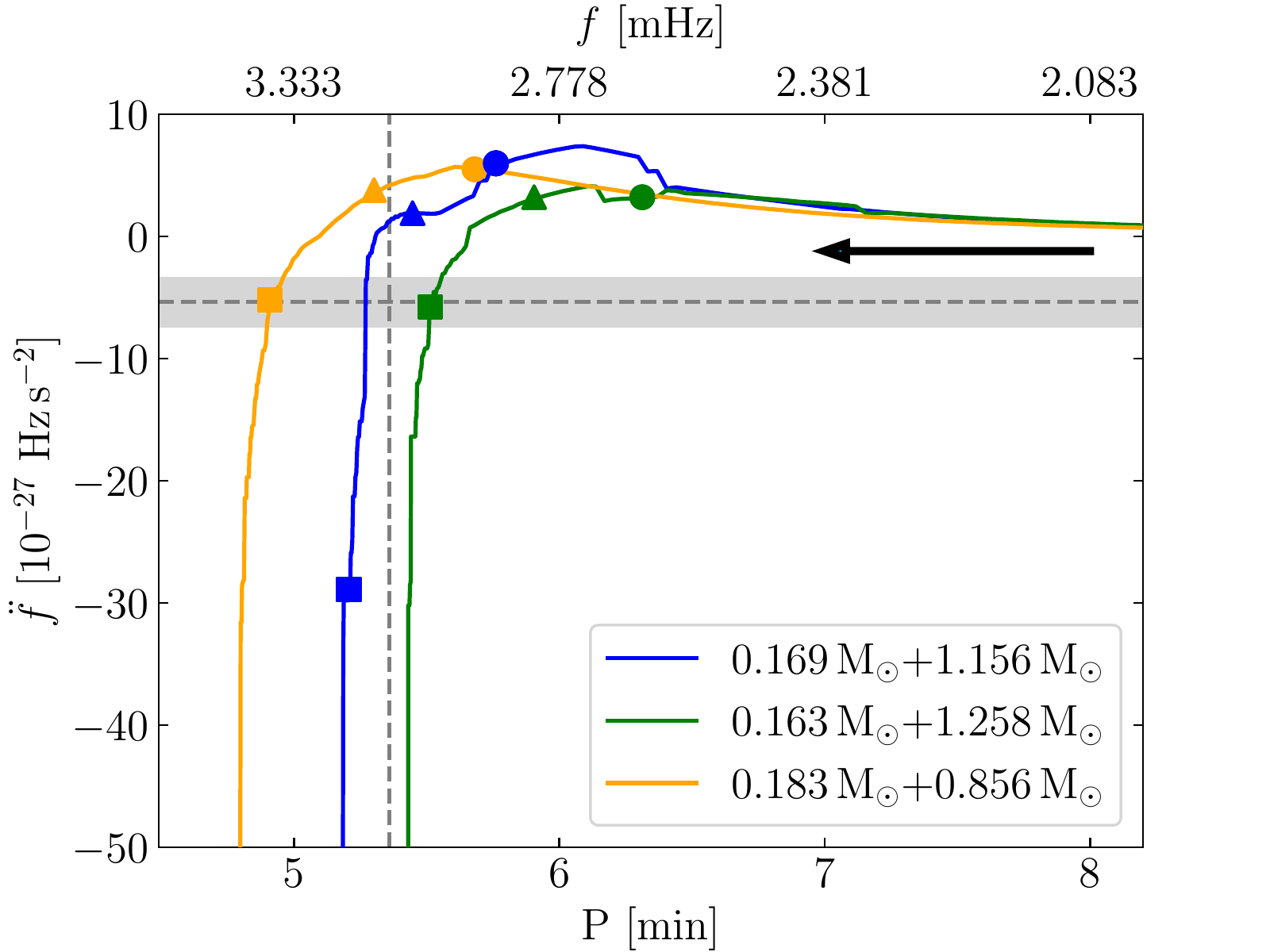}
    \caption{Models with evolved \citet[][]{Istrate2016} donor stars shown in the bottom right of Fig.~\ref{fig:MESAresults} for the mass transfer rate (left) and $\ddot f$ (right). Black arrows show the direction of evolution, grey dashed lines are the present state of the system, the shaded grey represents the error on $\ddot f_0$ and the time of H/He = 0.1, 0.05 and 0.01 is included with circles, triangles and squares respectively. Kinks in the mass transfer figure are again due to a direct-fed/disc-fed accretion transition.}
    \label{fig:MESAresultsmdot}
\end{figure*}
\subsection{MESA Results}
\label{subsec:MESAresults}
In the DWD configuration, the most significant variables influencing the time of maximum orbital frequency are the donor mass and the amount of hydrogen in the envelope \citep[see e.g.][]{Kaplan2012OrbitalEvolutionOfCompactWDBinaries}. The mass of the donor significantly affects the duration of Roche lobe overflow, whereas a larger hydrogen fraction induces a larger donor radius and earlier mass accretion. Besides the distinction between disc/direct fed accretion, combinations of accretor and donor masses control the magnitude of $\dot f$, where a larger accretor mass leads to a higher $\dot f$ for a given donor mass. Furthermore, the temperature of the donor star is decisive in controlling the onset of mass transfer and the timing of a negative $\ddot f$, since a higher temperature WD results in a larger radius.

For our chosen selection of hydrogen envelopes, we performed an in-depth search of the full parameter space of Fig.~\ref{fig:Allowed_masses} to determine which combinations of masses would lead to $\dot f\approx\dot f_0$ when $f = f_0$ at any stage of the system evolution before shortlisting to the best-agreeing candidate systems, those being where the sign of $\ddot f$ becomes negative near $f=f_0$. Our best-fitting models are shown in Fig.~\ref{fig:MESAresults}, whereby in all cases $\ddot f\lessapprox0$ at $f=f_0$.

We solely present the scenario where the donor star maintains an effective temperature of 5.5--6.0\,kK. This is at odds with the donor temperature derived in \cref{subsec:spectroscopy}, albeit a loose constraint. We attempted to investigate the scenario where the donor effective temperature is 16\,kK and found that it becomes extremely difficult to maintain any hydrogen envelope for all combinations of permitted masses (Fig.~\ref{fig:Allowed_masses}) in MESA. Increasing the donor temperature requires a more massive (and with this a smaller radius) donor for the binary to reach $f_0$ before $\ddot f \lesssim 0$ occurs. In the process, the binary reaches smaller orbital separations and we require a present-day mass transfer rate in excess of $10^{-6}$\,\(\textup{M}_\odot\)\,yr$^{-1}$, with the order of magnitude of $\ddot f$ to significantly increase over a few decades. Moreover, the total mass lost from a 16\,kK donor for tested models exceeded 0.015\,\(\textup{M}_\odot\) leading up to present day, such that an extremely thick hydrogen envelope before any mass transfer begins would be required for hydrogen to be observed in the spectrum. Our models therefore indicate a preference for cooler donor stars and smaller mass ratios.

We conclude from Fig.~\ref{fig:MESAresults} that, if we consider our full selection of hydrogen envelope masses, the donor star has a mass in the range $\approx$0.16--0.17\,\(\textup{M}_\odot\) to observe a negative $\ddot f$ at $f=f_0$. This result agrees with the expectation from \citet[][]{Kaplan2012OrbitalEvolutionOfCompactWDBinaries}, who suggest that the donor is an ELM WD. Additionally, we find that the accretor requires a minimum mass of 0.8$\,$\(\textup{M}_\odot\) to give rise to a sufficiently large $\dot f$ that matches our ephemeris with an ELM donor. These combinations of masses combined with the modelled separation indicate that the accretion stream would be close to the direct-disc mass transfer boundary, as is particularly noticeable by the kinks from the direct-to-disc accretion transition in Fig.~\ref{fig:MESAresults} for some models. For these boundary-models in particular, direct impact occurs in the last few 10,000s of years only, although the present mass transfer mechanism for all combinations of star masses presented involve direct impact\footnote{We note that \citet[][]{Barros2007HMCnc} state the condition that direct impact can only occur in HM~Cnc if 0.6~<~M$_1$~<~0.9$\,$\(\textup{M}_\odot\), however this was deduced with a zero-temperature masses-radius relationship. The masses we quote involve direct impact allowing for an accretor temperature T$_{\text{accretor}}=42\,500\,\text{K}$.}. Furthermore, all models show that the binary would not have deviated far from purely general-relativistic orbital decay, meaning that the true chirp mass of the system is comparable to the observed chirp mass of $0.3203\pm0.0001\,$\(\textup{M}_\odot\).

We use the 0.169\,\(\textup{M}_\odot\) + 1.156\,\(\textup{M}_\odot\) and 0.183\,\(\textup{M}_\odot\) + 0.856\,\(\textup{M}_\odot\) \citet[][]{Istrate2016} models presented in Figs.~\ref{fig:MESAresults} and \ref{fig:MESAresultsmdot} to check for consistency with the relative phase difference between the optical and X-ray peak flux. \citet[][]{Barros2007HMCnc} discuss how the X-ray signal arrives $\approx$0.26 cycles later than the optical. Our MESA models predict that the accretion stream directly impacts the accretor surface to produce a 0.35 and 0.30 cycle relative phase difference, respectively. This is approximately consistent with that observed, especially so as the accretor is likely to be at least partially inflated, acting to decrease the relative phase difference. For models with smaller initial hydrogen envelope masses, the relative phase differences are larger and show a worse accordance with the observations.

Our models show a consistent hydrogen envelope fraction with the $\text{H/He}=2.29\pm0.26$\% derived in \cref{subsec:spectroscopy}. We see from Figs.~\ref{fig:MESAresults} and \ref{fig:MESAresultsmdot} that all models have a hydrogen number abundance ratio of less than 10\% when $\ddot f = \ddot f_0$, and that for the more massive donor models presented the fraction is between 1\% and 5\%. While it may be possible for smaller starting mass hydrogen envelope models to match the hydrogen envelope fraction from our observations, we find a best agreement with the observations for higher starting envelope masses to be consistent with the location of the direct impact spot. A comparison of the full set of models does however show that a range of initial hydrogen envelope masses can produce DWD systems that reach the orbital period of HM~Cnc, such that the evolutionary channel is not exclusive to a very narrow set of initial conditions. Our models also show that the remaining hydrogen envelope of the donor will be almost entirely depleted by the time of maximum orbital frequency.

When we simulated a model similar to the 0.27$\,$\(\textup{M}_\odot\) donor and 0.55$\,$\(\textup{M}_\odot\) accretor predicted by \citet[][]{Roelofs2010HMCncMassRatio}, we found that the sign of $\ddot f$ would be $positive$ at the observed period for the full range of hydrogen envelopes. To emphasise, this would be the case for a cool donor, whereas hotter donors may be able to reproduce the observed $\dot f_0$ and $\ddot f_0$ at $f_0$, although the aforementioned complications when modelling a 16\,kK donor star make this difficult to explain. As well, analysis of $\ddot f$ indicates that it decreases much faster for a more massive donor, such that HM~Cnc would spend only centuries in a turn-around phase. Our models give a mass ratio $\approx$0.15--0.21, which disagrees with the mass ratio $0.27\leq q \leq 0.79$ inferred by \citet[][]{Roelofs2010HMCncMassRatio}. We suspect that the uncertainty in the origin of light of helium emission lines \citep[as discussed in][]{Roelofs2010HMCncMassRatio} and the choice of these lines to monitor the primary star makes it difficult to obtain absolute radial velocity measurements, so that the observational constraint on the mass ratio might be misleading even though a radial velocity variability is apparent.

As shown in Fig.~\ref{fig:MESAresultsmdot}, the magnitude of $\ddot f$ increases dramatically on approach to the maximum orbital frequency. This indicates that the timing of the frequency maximum will actually be in less than $2100\pm800$\,yrs and that deviations from a constant $\ddot f$ could become evident sooner than one might expect. The duration spent between orbital periods of 6--8 minutes is significantly longer, on the timescale of 100,000--200,000 years. It hence appears fortunate that HM~Cnc has been detected just before the maximum orbital frequency, which may be consequential of a strong observational bias. Though from Fig.~\ref{fig:MESAresultsmdot} we predict HM~Cnc's luminosity to increase upon approach to the maximum orbital frequency, with the mass transfer rate suspected to increase by at least 10$\times$, a clear observable difference is not expected in just 20 years (as suspected from \cref{subsec:Luminosity}).

The accretor MTRR radii in our best models for a 42\,500\,K accretor are 0.0055--0.0115\,\(\textup{R}_\odot\) for masses in the range 1.25--0.8\,\(\textup{M}_\odot\), respectively. In \cref{subsec:spectroscopy}, we explain that a radius constraint and a minimised flux-scaling-factor to our spectrum can be used to gain a distance measurement. Including the range of radii from our MESA models with this technique, we estimate a distance to HM~Cnc of $D$=1.05--2.24\,kpc.

We follow a similar methodology to \citet[][]{Strohmayer2021HMCnc} to compute the mass accretion rate with the observed X-ray luminosity. We assume that the accretion luminosity is equal to the difference between the gravitational energy at the first Lagrangian point, $\phi_{L1}$, and the surface of the accretor, $\phi_{R_{\textrm{a}}}$, such that the accretion luminosity L$_{\textrm{acc}} = \dot M \left(\phi_{L1}-\phi_{R_{\textrm{a}}}\right)$,  with $\phi_{L1}$ and $\phi_{R_{\textrm{a}}}$ calculable with the prescription of \citet[][]{HanAndWebbink1999}. Taking donor and accretor masses of 0.17\,\(\textup{M}_\odot\) and 1.0\,\(\textup{M}_\odot\) with an accretor radius of 0.01\,\(\textup{R}_\odot\) and the X-ray luminosity to be  1.8$\times10^{33}(\Delta\Omega/4\pi)D_{\textrm{kpc}}^2$\,erg\,s$^{-1}$ \citep[][]{Strohmayer2021HMCnc}, we find a mass transfer rate $\dot M=1.84\times10^{-10}\times D_\textrm{kpc}^2$\,\(\textup{M}_\odot\)\,yr$^{-1}$. If we assume a distance of 2\,kpc, $\dot M=10^{-9.1}$\,\(\textup{M}_\odot\)\,yr$^{-1}$. As recognisable by Fig.~\ref{fig:MESAresultsmdot}, this would be a clear pitfall of our models, where we would be expecting $\dot M\approx10^{-7.5}$\,\(\textup{M}_\odot\)\,yr$^{-1}$. If the distance was instead 5\,kpc, $\dot M =10^{-8.3}$\,\(\textup{M}_\odot\)\,yr$^{-1}$, such that assuming larger distances to the source would make our models consistent. An exception is that there could be a missing luminosity component in the extreme ultra-violet that we are not sensitive to with the X-ray nor our own observations, meaning that the X-ray luminosity under-predicts the mass accretion rate. For both the 5.5--6\,kK and the 16\,kK WD donors, and with the full grid of hydrogen models, we struggle to find a $\dot M$ consistent with 10$^{-9}-10^{-8}$\,\(\textup{M}_\odot\)\,yr$^{-1}$ that would generate a negative $\ddot f_0$, since the higher mass accretion rates are associated with the timing of the turn-around phase.

As evident, there are still components of HM~Cnc that remain unknown which make a precise characterisation of the system difficult. However, our models do show a good agreement with multiple observed properties, including the hydrogen fraction, the location of the impact spot and all components of the cubic ephemeris. Our models favour larger initial hydrogen envelope masses and are consistent with the theory outlined by \citet[][]{DAntona2006} as an origin for HM~Cnc and the ELM donor conclusion of \citet[][]{Kaplan2012OrbitalEvolutionOfCompactWDBinaries}. Our models show that it is unlikely that HM~Cnc had a hydrogen envelope thinner than 10$^{-3}$\,\(\textup{M}_\odot\) before the onset of mass transfer and require thicker initial hydrogen envelopes to be consistent with the location of the accretion impact spot. Better quantification of the donor temperature and an accurate distance measurement clearly remain the key to settle the nature of HM~Cnc. If $D\approx2$\,kpc, our models would need to be tweaked to include a hotter donor temperature, a more massive initial hydrogen envelope or a mixture of the two to agree with the $f_0$, $\dot f_0$ and $\ddot f_0$ presented in this paper, since these components both permit larger mass donor stars and smaller mass accretors with an increased mass transfer rate.

\section{Conclusions}
\label{sec:Conclusion}
We presented new timing measurements for HM~Cnc following an optical observing campaign to monitor the binary over the last 20 years. We measured $\dot f_0$ precise to the 0.03\% level while obtaining $\ddot f_0~=~(-5.38\pm2.10)~\times 10^{-27}$~Hz\,s$^{-2}$. Our ephemeris affirms that the orbital frequency is still increasing, however the negative $\ddot f_0$ implies that HM~Cnc may reach a maximum frequency within $2100\pm800$\,yrs from now.

Then, we outlined a method whereby knowing $\ddot f_0$ provides a tight condition on the present state of the system. We used MESA to explore the DWD AM~CVn configuration. Since we directly witness the presence of hydrogen in our HST spectra, we surveyed multiple masses of donor hydrogen envelopes with combinations of permitted donor and accretor masses. Our best model for a cool and initially hydrogen-rich donor indicates system masses of M$_{\textrm{d}}\approx0.17$\,\(\textup{M}_\odot\) and M$_{\textrm{a}}\approx1.0$\,\(\textup{M}_\odot\). Our model is able to reproduce the observed ephemeris, the hydrogen envelope fraction and the location of the direct impact spot, though suffers from finding a mass transfer rate higher than suggested from X-ray observations if a distance of less than 5\,kpc is to be assumed. Furthermore, our model indicates that, after reaching a maximum frequency, the orbital frequency will decrease as the orbital separation increases, such that the binary is unlikely to merge unless an imminent helium-powered nova event \citep[][]{Shen2015EveryDWDmerge} or a sub-Chandrasekhar mass type \Romannum{1}a supernova \citep[e.g.][]{Shen2018SubChandrasekhar1a} is to occur.

Our MESA models predict that HM~Cnc's $\dot f_0$ is close to that expected from the purely general relativistic orbital decay for two point masses. This means that the observed system chirp mass ($0.3203\pm0.0001\,$\(\textup{M}_\odot\)) is reflective of the true system chirp mass (0.3347\,\(\textup{M}_\odot\) for a 0.17\,\(\textup{M}_\odot\) + 1.0\,\(\textup{M}_\odot\) combination) and predicts a characteristic strain of 2.67\,$\times$\,10$^{-19}$/$D_{\textrm{kpc}}$ after a 4 year observation time with LISA. Furthermore, a 0.17$\,$\(\textup{M}_\odot\)~+~1.0$\,$\(\textup{M}_\odot\) star mass pairing, an inclination of \ang{\approx 38} \citep[][although this could differ if the radial velocities are misleading, see \cref{subsec:MESAresults}]{Roelofs2010HMCncMassRatio} and a gravitational wave polarisation angle of \ang{0} generate a LISA SNR of 147 after a 4 year mission for a distance of 5\,kpc\footnote{\text{https://heasarc.gsfc.nasa.gov/lisa/lisatool/}}. Even at 5\,kpc, HM~Cnc will be one of the highest SNR binary star systems detectable by the LISA spacecraft and is still an ideal reference source for the TianQin spacecraft.

\section*{Acknowledgements}
We thank the referee for their helpful report. Based on observations made with the NASA/ESA Hubble Space Telescope, obtained at the Space Telescope Science Institute, which is operated by the Association of Universities for Research in Astronomy, Inc., under NASA contract NAS 5-26555. The associated proposal for this program is \#10806. Based on observations made with the Gran Telescopio Canarias (GTC), installed at the Spanish Observatorio del Roque de los Muchachos of the Instituto de Astrofísica de Canarias, on the island of La Palma. Based on observations made with the Nordic Optical Telescope, owned in collaboration by the University of Turku and Aarhus University, and operated jointly by Aarhus University, the University of Turku and the University of Oslo, representing Denmark, Finland and Norway, the University of Iceland and Stockholm University at the Observatorio del Roque de los Muchachos, La Palma, Spain, of the Instituto de Astrofisica de Canarias. The data presented here were obtained (in part) with ALFOSC, which is provided by the Instituto de Astrofisica de Andalucia (IAA) under a joint agreement with the University of Copenhagen and NOT. This paper includes observations made in the Observatorios de Canarias del IAC with the WHT and INT operated on the island of La Palma by the IAC in the Observatorio del Roque de los Muchachos. The design and construction of HiPERCAM was funded by the European Research Council under the European Union's Seventh Framework Programme (FP/2007-2013) under ERC-2013-ADG Grant Agreement no. 340040 (HiPERCAM). ULTRASPEC, ULTRACAM and HiPERCAM operations, and VSD, are funded by the Science and Technology Facilities Council (grant ST/V000853/1). JM was supported by funding from a Science and Technology Facilities Council (STFC) studentship. TRM and IP were supported by STFC grant ST/T000406/1. DS acknowledges support from the STFC via grants ST/T007184/1, ST/T003103/1 and ST/T000406/1. SGP acknowledges the support of a STFC Ernest Rutherford Fellowship. For the purpose of open access, the authors has applied a creative commons attribution (CC BY) licence to any author accepted manuscript version arising. JM thanks members of the MESA mailing list and the Astrometry.net discussion forum for their helpful advice. This research made use of Astropy \citep[][]{astropy:2013,astropy:2018}, NumPy \citep{Harris2020Numpy} and SciPy \citep[][]{2020SciPy-NMeth} routines. 

\section*{Data Availability}
Timing solutions are included in the Appendix for use. The raw photometric data and reduced HST spectra will be made available upon reasonable request to the authors. The digitised X-ray timing will be made available upon request if necessary.



\bibliographystyle{mnras}
\bibliography{HMCnc} 

\begin{thebibliography}{}
\makeatletter
\relax
\def\mn@urlcharsother{\let\do\@makeother \do\$\do\&\do\#\do\^\do\_\do\%\do\~}
\def\mn@doi{\begingroup\mn@urlcharsother \@ifnextchar [ {\mn@doi@}
  {\mn@doi@[]}}
\def\mn@doi@[#1]#2{\def\@tempa{#1}\ifx\@tempa\@empty \href
  {http://dx.doi.org/#2} {doi:#2}\else \href {http://dx.doi.org/#2} {#1}\fi
  \endgroup}
\def\mn@eprint#1#2{\mn@eprint@#1:#2::\@nil}
\def\mn@eprint@arXiv#1{\href {http://arxiv.org/abs/#1} {{\tt arXiv:#1}}}
\def\mn@eprint@dblp#1{\href {http://dblp.uni-trier.de/rec/bibtex/#1.xml}
  {dblp:#1}}
\def\mn@eprint@#1:#2:#3:#4\@nil{\def\@tempa {#1}\def\@tempb {#2}\def\@tempc
  {#3}\ifx \@tempc \@empty \let \@tempc \@tempb \let \@tempb \@tempa \fi \ifx
  \@tempb \@empty \def\@tempb {arXiv}\fi \@ifundefined
  {mn@eprint@\@tempb}{\@tempb:\@tempc}{\expandafter \expandafter \csname
  mn@eprint@\@tempb\endcsname \expandafter{\@tempc}}}

\bibitem[\protect\citeauthoryear{{Amaro-Seoane} et~al.,}{{Amaro-Seoane}
  et~al.}{2017}]{LISA}
{Amaro-Seoane} P.,  et~al., 2017, arXiv e-prints, \href
  {https://ui.adsabs.harvard.edu/abs/2017arXiv170200786A} {p. arXiv:1702.00786}

\bibitem[\protect\citeauthoryear{{Amaro-Seoane} et~al.,}{{Amaro-Seoane}
  et~al.}{2022}]{AmaroSeoane2022LISAastrophysics}
{Amaro-Seoane} P.,  et~al., 2022, arXiv e-prints, \href
  {https://ui.adsabs.harvard.edu/abs/2022arXiv220306016A} {p. arXiv:2203.06016}

\bibitem[\protect\citeauthoryear{{Appenzeller} et~al.,}{{Appenzeller}
  et~al.}{1998}]{VLTFors1998}
{Appenzeller} I.,  et~al., 1998, The Messenger, \href
  {https://ui.adsabs.harvard.edu/abs/1998Msngr..94....1A} {94, 1}

\bibitem[\protect\citeauthoryear{{Astropy Collaboration} et~al.,}{{Astropy
  Collaboration} et~al.}{2013}]{astropy:2013}
{Astropy Collaboration} et~al., 2013, \mn@doi [\aap]
  {10.1051/0004-6361/201322068}, \href
  {http://adsabs.harvard.edu/abs/2013A%26A...558A..33A} {558, A33}

\bibitem[\protect\citeauthoryear{{Astropy Collaboration} et~al.,}{{Astropy
  Collaboration} et~al.}{2018}]{astropy:2018}
{Astropy Collaboration} et~al., 2018, \mn@doi [\aj] {10.3847/1538-3881/aabc4f},
  \href {https://ui.adsabs.harvard.edu/abs/2018AJ....156..123A} {156, 123}

\bibitem[\protect\citeauthoryear{{Barros} et~al.,}{{Barros}
  et~al.}{2007}]{Barros2007HMCnc}
{Barros} S.~C.~C.,  et~al., 2007, \mn@doi [\mnras]
  {10.1111/j.1365-2966.2006.11244.x}, \href
  {https://ui.adsabs.harvard.edu/abs/2007MNRAS.374.1334B} {374, 1334}

\bibitem[\protect\citeauthoryear{{B{\'e}dard}, {Bergeron}, {Brassard}  \&
  {Fontaine}}{{B{\'e}dard} et~al.}{2020}]{Bedard2020MontrealWDModelsMTR}
{B{\'e}dard} A.,  {Bergeron} P.,  {Brassard} P.,   {Fontaine} G.,  2020,
  \mn@doi [\apj] {10.3847/1538-4357/abafbe}, \href
  {https://ui.adsabs.harvard.edu/abs/2020ApJ...901...93B} {901, 93}

\bibitem[\protect\citeauthoryear{{Buzzoni} et~al.,}{{Buzzoni}
  et~al.}{1984}]{ESOEFOSC2}
{Buzzoni} B.,  et~al., 1984, The Messenger, \href
  {https://ui.adsabs.harvard.edu/abs/1984Msngr..38....9B} {38, 9}

\bibitem[\protect\citeauthoryear{{Chen}, {Chen}  \& {Han}}{{Chen}
  et~al.}{2022}]{Chen2022MESAamcvnWDdonor}
{Chen} H.-L.,  {Chen} X.,   {Han} Z.,  2022, arXiv e-prints, \href
  {https://ui.adsabs.harvard.edu/abs/2022arXiv220704592C} {p. arXiv:2207.04592}

\bibitem[\protect\citeauthoryear{{Cheng}, {Cummings}  \& {M{\'e}nard}}{{Cheng}
  et~al.}{2019}]{Cheng2019CoolingAndTransverseVel}
{Cheng} S.,  {Cummings} J.~D.,   {M{\'e}nard} B.,  2019, \mn@doi [\apj]
  {10.3847/1538-4357/ab4989}, \href
  {https://ui.adsabs.harvard.edu/abs/2019ApJ...886..100C} {886, 100}

\bibitem[\protect\citeauthoryear{{D'Antona}, {Ventura}, {Burderi}  \&
  {Teodorescu}}{{D'Antona} et~al.}{2006}]{DAntona2006}
{D'Antona} F.,  {Ventura} P.,  {Burderi} L.,   {Teodorescu} A.,  2006, \mn@doi
  [\apj] {10.1086/507408}, \href
  {https://ui.adsabs.harvard.edu/abs/2006ApJ...653.1429D} {653, 1429}

\bibitem[\protect\citeauthoryear{{Dall'Osso}, {Israel}  \&
  {Stella}}{{Dall'Osso} et~al.}{2007}]{DallOsso2007UIHMCnc}
{Dall'Osso} S.,  {Israel} G.~L.,   {Stella} L.,  2007, \mn@doi [\aap]
  {10.1051/0004-6361:20064862}, \href
  {https://ui.adsabs.harvard.edu/abs/2007A&A...464..417D} {464, 417}

\bibitem[\protect\citeauthoryear{{Deloye} \& {Taam}}{{Deloye} \&
  {Taam}}{2006}]{Deloye2006predictingFddotFromLacc}
{Deloye} C.~J.,  {Taam} R.~E.,  2006, \mn@doi [\apjl] {10.1086/508372}, \href
  {https://ui.adsabs.harvard.edu/abs/2006ApJ...649L..99D} {649, L99}

\bibitem[\protect\citeauthoryear{{Dhillon} et~al.,}{{Dhillon}
  et~al.}{2007}]{ULTRACAM2007}
{Dhillon} V.~S.,  et~al., 2007, \mn@doi [\mnras]
  {10.1111/j.1365-2966.2007.11881.x}, \href
  {https://ui.adsabs.harvard.edu/abs/2007MNRAS.378..825D} {378, 825}

\bibitem[\protect\citeauthoryear{{Dhillon} et~al.,}{{Dhillon}
  et~al.}{2014}]{ULTRASPEC2014}
{Dhillon} V.~S.,  et~al., 2014, \mn@doi [\mnras] {10.1093/mnras/stu1660}, \href
  {https://ui.adsabs.harvard.edu/abs/2014MNRAS.444.4009D} {444, 4009}

\bibitem[\protect\citeauthoryear{{Dhillon} et~al.,}{{Dhillon}
  et~al.}{2016}]{HiPERCAM2016}
{Dhillon} V.~S.,  et~al., 2016, in {Evans} C.~J.,  {Simard} L.,   {Takami} H.,
  eds,  Society of Photo-Optical Instrumentation Engineers (SPIE) Conference
  Series Vol. 9908, Ground-based and Airborne Instrumentation for Astronomy VI.
  p. 99080Y (\mn@eprint {arXiv} {1606.09214}), \mn@doi{10.1117/12.2229055}

\bibitem[\protect\citeauthoryear{{Dhillon} et~al.,}{{Dhillon}
  et~al.}{2021}]{Hipercam2021Paper}
{Dhillon} V.~S.,  et~al., 2021, \mn@doi [\mnras] {10.1093/mnras/stab2130},
  \href {https://ui.adsabs.harvard.edu/abs/2021MNRAS.507..350D} {507, 350}

\bibitem[\protect\citeauthoryear{{Eggleton}}{{Eggleton}}{1983}]{Eggleton1983}
{Eggleton} P.~P.,  1983, \mn@doi [\apj] {10.1086/160960}, \href
  {https://ui.adsabs.harvard.edu/abs/1983ApJ...268..368E} {268, 368}

\bibitem[\protect\citeauthoryear{{Esposito}, {Israel}, {Dall'Osso}  \&
  {Covino}}{{Esposito} et~al.}{2014}]{Esposito2014HMCncSWIFT}
{Esposito} P.,  {Israel} G.~L.,  {Dall'Osso} S.,   {Covino} S.,  2014, \mn@doi
  [\aap] {10.1051/0004-6361/201322719}, \href
  {https://ui.adsabs.harvard.edu/abs/2014A&A...561A.117E} {561, A117}

\bibitem[\protect\citeauthoryear{{Fitzpatrick}}{{Fitzpatrick}}{1999}]{Fitzpatrick99}
{Fitzpatrick} E.~L.,  1999, \mn@doi [\pasp] {10.1086/316293}, \href
  {https://ui.adsabs.harvard.edu/abs/1999PASP..111...63F} {111, 63}

\bibitem[\protect\citeauthoryear{{Gaia Collaboration} et~al.,}{{Gaia
  Collaboration} et~al.}{2016}]{GaiaMissionPaper}
{Gaia Collaboration} et~al., 2016, \mn@doi [\aap]
  {10.1051/0004-6361/201629272}, \href
  {https://ui.adsabs.harvard.edu/abs/2016A&A...595A...1G} {595, A1}

\bibitem[\protect\citeauthoryear{{Hakala}, {Ramsay}, {Wu}, {Hjalmarsdotter},
  {J{\"a}rvinen}, {J{\"a}rvinen}  \& {Cropper}}{{Hakala}
  et~al.}{2003}]{Hakala2003HMCncBinary}
{Hakala} P.,  {Ramsay} G.,  {Wu} K.,  {Hjalmarsdotter} L.,  {J{\"a}rvinen} S.,
  {J{\"a}rvinen} A.,   {Cropper} M.,  2003, \mn@doi [\mnras]
  {10.1046/j.1365-8711.2003.06830.x}, \href
  {https://ui.adsabs.harvard.edu/abs/2003MNRAS.343L..10H} {343, L10}

\bibitem[\protect\citeauthoryear{{Hakala}, {Ramsay}  \& {Byckling}}{{Hakala}
  et~al.}{2004}]{Hakala2004HMCncmore}
{Hakala} P.,  {Ramsay} G.,   {Byckling} K.,  2004, \mn@doi [\mnras]
  {10.1111/j.1365-2966.2004.08074.x}, \href
  {https://ui.adsabs.harvard.edu/abs/2004MNRAS.353..453H} {353, 453}

\bibitem[\protect\citeauthoryear{{Han} \& {Webbink}}{{Han} \&
  {Webbink}}{1999}]{HanAndWebbink1999}
{Han} Z.,  {Webbink} R.~F.,  1999, \aap, \href
  {https://ui.adsabs.harvard.edu/abs/1999A&A...349L..17H} {349, L17}

\bibitem[\protect\citeauthoryear{{Harris} et~al.,}{{Harris}
  et~al.}{2020}]{Harris2020Numpy}
{Harris} C.~R.,  et~al., 2020, \mn@doi [\nat] {10.1038/s41586-020-2649-2},
  \href {https://ui.adsabs.harvard.edu/abs/2020Natur.585..357H} {585, 357}

\bibitem[\protect\citeauthoryear{{Hurley}, {Tout}  \& {Pols}}{{Hurley}
  et~al.}{2002}]{Hurley2002BinaryEvolution}
{Hurley} J.~R.,  {Tout} C.~A.,   {Pols} O.~R.,  2002, \mn@doi [\mnras]
  {10.1046/j.1365-8711.2002.05038.x}, \href
  {https://ui.adsabs.harvard.edu/abs/2002MNRAS.329..897H} {329, 897}

\bibitem[\protect\citeauthoryear{{Israel} et~al.,}{{Israel}
  et~al.}{2002}]{Israel2002HMCncDiscovery}
{Israel} G.~L.,  et~al., 2002, \mn@doi [\aap] {10.1051/0004-6361:20020314},
  \href {https://ui.adsabs.harvard.edu/abs/2002A&A...386L..13I} {386, L13}

\bibitem[\protect\citeauthoryear{{Israel} et~al.,}{{Israel}
  et~al.}{2004}]{Israel2004HMCncPulse}
{Israel} G.~L.,  et~al., 2004, Memorie della Societa Astronomica Italiana
  Supplementi, \href {https://ui.adsabs.harvard.edu/abs/2004MSAIS...5..148I}
  {5, 148}

\bibitem[\protect\citeauthoryear{{Istrate}, {Marchant}, {Tauris}, {Langer},
  {Stancliffe}  \& {Grassitelli}}{{Istrate} et~al.}{2016}]{Istrate2016}
{Istrate} A.~G.,  {Marchant} P.,  {Tauris} T.~M.,  {Langer} N.,  {Stancliffe}
  R.~J.,   {Grassitelli} L.,  2016, \mn@doi [\aap]
  {10.1051/0004-6361/201628874}, \href
  {https://ui.adsabs.harvard.edu/abs/2016A&A...595A..35I} {595, A35}

\bibitem[\protect\citeauthoryear{{Kaplan}, {Bildsten}  \& {Steinfadt}}{{Kaplan}
  et~al.}{2012}]{Kaplan2012OrbitalEvolutionOfCompactWDBinaries}
{Kaplan} D.~L.,  {Bildsten} L.,   {Steinfadt} J. D.~R.,  2012, \mn@doi [\apj]
  {10.1088/0004-637X/758/1/64}, \href
  {https://ui.adsabs.harvard.edu/abs/2012ApJ...758...64K} {758, 64}

\bibitem[\protect\citeauthoryear{{Kim}, {L{\'e}pine}  \& {Medan}}{{Kim}
  et~al.}{2020}]{Kim2020HaloVelocities}
{Kim} B.,  {L{\'e}pine} S.,   {Medan} I.,  2020, \mn@doi [\apj]
  {10.3847/1538-4357/aba523}, \href
  {https://ui.adsabs.harvard.edu/abs/2020ApJ...899...83K} {899, 83}

\bibitem[\protect\citeauthoryear{{Koester}}{{Koester}}{2010}]{Koester2010WDmodels}
{Koester} D.,  2010, \memsai, \href
  {https://ui.adsabs.harvard.edu/abs/2010MmSAI..81..921K} {81, 921}

\bibitem[\protect\citeauthoryear{{Kupfer} et~al.,}{{Kupfer}
  et~al.}{2018}]{Kupfer2018LISAverificationBinaries}
{Kupfer} T.,  et~al., 2018, \mn@doi [\mnras] {10.1093/mnras/sty1545}, \href
  {https://ui.adsabs.harvard.edu/abs/2018MNRAS.480..302K} {480, 302}

\bibitem[\protect\citeauthoryear{{Lang}, {Hogg}, {Mierle}, {Blanton}  \&
  {Roweis}}{{Lang} et~al.}{2010}]{Astrometrynet2010}
{Lang} D.,  {Hogg} D.~W.,  {Mierle} K.,  {Blanton} M.,   {Roweis} S.,  2010,
  \mn@doi [\aj] {10.1088/0004-6256/139/5/1782}, \href
  {https://ui.adsabs.harvard.edu/abs/2010AJ....139.1782L} {139, 1782}

\bibitem[\protect\citeauthoryear{{Luo} et~al.,}{{Luo}
  et~al.}{2016}]{TianQinProposal2016}
{Luo} J.,  et~al., 2016, \mn@doi [Classical and Quantum Gravity]
  {10.1088/0264-9381/33/3/035010}, \href
  {https://ui.adsabs.harvard.edu/abs/2016CQGra..33c5010L} {33, 035010}

\bibitem[\protect\citeauthoryear{{Maoz}, {Mannucci}  \& {Nelemans}}{{Maoz}
  et~al.}{2014}]{Maoz2014Type1aProgenitors}
{Maoz} D.,  {Mannucci} F.,   {Nelemans} G.,  2014, \mn@doi [\araa]
  {10.1146/annurev-astro-082812-141031}, \href
  {https://ui.adsabs.harvard.edu/abs/2014ARA&A..52..107M} {52, 107}

\bibitem[\protect\citeauthoryear{{Marsh}, {Nelemans}  \& {Steeghs}}{{Marsh}
  et~al.}{2004}]{Marsh2004WDMassTransfer}
{Marsh} T.~R.,  {Nelemans} G.,   {Steeghs} D.,  2004, \mn@doi [\mnras]
  {10.1111/j.1365-2966.2004.07564.x}, \href
  {https://ui.adsabs.harvard.edu/abs/2004MNRAS.350..113M} {350, 113}

\bibitem[\protect\citeauthoryear{{Mason} et~al.,}{{Mason}
  et~al.}{2010}]{Mason2010HMcncSpectraVLT}
{Mason} E.,  et~al., 2010, arXiv e-prints, \href
  {https://ui.adsabs.harvard.edu/abs/2010arXiv1003.1986M} {p. arXiv:1003.1986}

\bibitem[\protect\citeauthoryear{{Miller Bertolami}, {Battich}, {C{\'o}rsico},
  {Althaus}  \& {Wachlin}}{{Miller Bertolami}
  et~al.}{2022}]{Bertolami2022SubdwarfFromCOWDplusHEWD}
{Miller Bertolami} M.~M.,  {Battich} T.,  {C{\'o}rsico} A.~H.,  {Althaus}
  L.~G.,   {Wachlin} F.~C.,  2022, \mn@doi [\mnras] {10.1093/mnrasl/slab134},
  \href {https://ui.adsabs.harvard.edu/abs/2022MNRAS.511L..60M} {511, L60}

\bibitem[\protect\citeauthoryear{{Naylor}}{{Naylor}}{1998}]{OptimalPhotometry1998Naylor}
{Naylor} T.,  1998, \mn@doi [\mnras] {10.1046/j.1365-8711.1998.01314.x}, \href
  {https://ui.adsabs.harvard.edu/abs/1998MNRAS.296..339N} {296, 339}

\bibitem[\protect\citeauthoryear{{Nelemans}, {Yungelson}, {Portegies Zwart}  \&
  {Verbunt}}{{Nelemans} et~al.}{2001a}]{NelemansI2001populationSynthesisOfWDs}
{Nelemans} G.,  {Yungelson} L.~R.,  {Portegies Zwart} S.~F.,   {Verbunt} F.,
  2001a, \mn@doi [\aap] {10.1051/0004-6361:20000147}, \href
  {https://ui.adsabs.harvard.edu/abs/2001A&A...365..491N} {365, 491}

\bibitem[\protect\citeauthoryear{{Nelemans}, {Portegies Zwart}, {Verbunt}  \&
  {Yungelson}}{{Nelemans}
  et~al.}{2001b}]{NelemansII2001populationSynthesisOfWDsAMCVn}
{Nelemans} G.,  {Portegies Zwart} S.~F.,  {Verbunt} F.,   {Yungelson} L.~R.,
  2001b, \mn@doi [\aap] {10.1051/0004-6361:20010049}, \href
  {https://ui.adsabs.harvard.edu/abs/2001A&A...368..939N} {368, 939}

\bibitem[\protect\citeauthoryear{{Norton}, {Haswell}  \& {Wynn}}{{Norton}
  et~al.}{2004}]{Norton2004IPmodelHMCnc}
{Norton} A.~J.,  {Haswell} C.~A.,   {Wynn} G.~A.,  2004, \mn@doi [\aap]
  {10.1051/0004-6361:20035906}, \href
  {https://ui.adsabs.harvard.edu/abs/2004A&A...419.1025N} {419, 1025}

\bibitem[\protect\citeauthoryear{{Pauli}, {Napiwotzki}, {Heber}, {Altmann}  \&
  {Odenkirchen}}{{Pauli} et~al.}{2006}]{Pauli20063Dkinematics}
{Pauli} E.~M.,  {Napiwotzki} R.,  {Heber} U.,  {Altmann} M.,   {Odenkirchen}
  M.,  2006, \mn@doi [\aap] {10.1051/0004-6361:20052730}, \href
  {https://ui.adsabs.harvard.edu/abs/2006A&A...447..173P} {447, 173}

\bibitem[\protect\citeauthoryear{{Paxton}, {Bildsten}, {Dotter}, {Herwig},
  {Lesaffre}  \& {Timmes}}{{Paxton} et~al.}{2011}]{Mesa1}
{Paxton} B.,  {Bildsten} L.,  {Dotter} A.,  {Herwig} F.,  {Lesaffre} P.,
  {Timmes} F.,  2011, \mn@doi [\apjs] {10.1088/0067-0049/192/1/3}, \href
  {https://ui.adsabs.harvard.edu/abs/2011ApJS..192....3P} {192, 3}

\bibitem[\protect\citeauthoryear{{Paxton} et~al.,}{{Paxton}
  et~al.}{2013}]{Mesa2}
{Paxton} B.,  et~al., 2013, \mn@doi [\apjs] {10.1088/0067-0049/208/1/4}, \href
  {https://ui.adsabs.harvard.edu/abs/2013ApJS..208....4P} {208, 4}

\bibitem[\protect\citeauthoryear{{Paxton} et~al.,}{{Paxton}
  et~al.}{2015}]{Mesa3binary}
{Paxton} B.,  et~al., 2015, \mn@doi [\apjs] {10.1088/0067-0049/220/1/15}, \href
  {https://ui.adsabs.harvard.edu/abs/2015ApJS..220...15P} {220, 15}

\bibitem[\protect\citeauthoryear{{Paxton} et~al.,}{{Paxton}
  et~al.}{2018}]{Mesa4}
{Paxton} B.,  et~al., 2018, \mn@doi [\apjs] {10.3847/1538-4365/aaa5a8}, \href
  {https://ui.adsabs.harvard.edu/abs/2018ApJS..234...34P} {234, 34}

\bibitem[\protect\citeauthoryear{{Paxton} et~al.,}{{Paxton}
  et~al.}{2019}]{Mesa5}
{Paxton} B.,  et~al., 2019, \mn@doi [\apjs] {10.3847/1538-4365/ab2241}, \href
  {https://ui.adsabs.harvard.edu/abs/2019ApJS..243...10P} {243, 10}

\bibitem[\protect\citeauthoryear{{Pelisoli} et~al.,}{{Pelisoli}
  et~al.}{2021}]{Pelisoli2021FastSpinningWD}
{Pelisoli} I.,  et~al., 2021, \mn@doi [\mnras] {10.1093/mnras/stab2511}, \href
  {https://ui.adsabs.harvard.edu/abs/2021MNRAS.507.6132P} {507, 6132}

\bibitem[\protect\citeauthoryear{Price-Whelan}{Price-Whelan}{2017}]{gala}
Price-Whelan A.~M.,  2017, \mn@doi [The Journal of Open Source Software]
  {10.21105/joss.00388}, 2

\bibitem[\protect\citeauthoryear{{Ramsay} \& {Hakala}}{{Ramsay} \&
  {Hakala}}{2005}]{RATShmCnc2005}
{Ramsay} G.,  {Hakala} P.,  2005, \mn@doi [\mnras]
  {10.1111/j.1365-2966.2005.09035.x}, \href
  {https://ui.adsabs.harvard.edu/abs/2005MNRAS.360..314R} {360, 314}

\bibitem[\protect\citeauthoryear{{Ramsay}, {Hakala}  \& {Cropper}}{{Ramsay}
  et~al.}{2002}]{RamsayHakalaCropper2002HMCncBinarydiscoveryMaybe}
{Ramsay} G.,  {Hakala} P.,   {Cropper} M.,  2002, \mn@doi [\mnras]
  {10.1046/j.1365-8711.2002.05471.x}, \href
  {https://ui.adsabs.harvard.edu/abs/2002MNRAS.332L...7R} {332, L7}

\bibitem[\protect\citeauthoryear{{Reinsch}, {Steiper}  \& {Dreizler}}{{Reinsch}
  et~al.}{2007}]{Reinsch2007HMCncFluxes}
{Reinsch} K.,  {Steiper} J.,   {Dreizler} S.,  2007, in {Napiwotzki} R.,
  {Burleigh} M.~R.,  eds,  Astronomical Society of the Pacific Conference
  Series Vol. 372, 15th European Workshop on White Dwarfs. p.~419

\bibitem[\protect\citeauthoryear{{Ritter}}{{Ritter}}{1988}]{Ritter1988}
{Ritter} H.,  1988, \aap, \href
  {https://ui.adsabs.harvard.edu/abs/1988A%26A...202...93R} {202, 93}

\bibitem[\protect\citeauthoryear{{Roelofs}, {Rau}, {Marsh}, {Steeghs}, {Groot}
  \& {Nelemans}}{{Roelofs} et~al.}{2010}]{Roelofs2010HMCncMassRatio}
{Roelofs} G. H.~A.,  {Rau} A.,  {Marsh} T.~R.,  {Steeghs} D.,  {Groot} P.~J.,
  {Nelemans} G.,  2010, \mn@doi [\apjl] {10.1088/2041-8205/711/2/L138}, \href
  {https://ui.adsabs.harvard.edu/abs/2010ApJ...711L.138R} {711, L138}

\bibitem[\protect\citeauthoryear{{Shen}}{{Shen}}{2015}]{Shen2015EveryDWDmerge}
{Shen} K.~J.,  2015, \mn@doi [\apjl] {10.1088/2041-8205/805/1/L6}, \href
  {https://ui.adsabs.harvard.edu/abs/2015ApJ...805L...6S} {805, L6}

\bibitem[\protect\citeauthoryear{{Shen}, {Kasen}, {Miles}  \&
  {Townsley}}{{Shen} et~al.}{2018}]{Shen2018SubChandrasekhar1a}
{Shen} K.~J.,  {Kasen} D.,  {Miles} B.~J.,   {Townsley} D.~M.,  2018, \mn@doi
  [\apj] {10.3847/1538-4357/aaa8de}, \href
  {https://ui.adsabs.harvard.edu/abs/2018ApJ...854...52S} {854, 52}

\bibitem[\protect\citeauthoryear{{Shklovskii}}{{Shklovskii}}{1970}]{Shklovskii1970effect}
{Shklovskii} I.~S.,  1970, \sovast, \href
  {https://ui.adsabs.harvard.edu/abs/1970SvA....13..562S} {13, 562}

\bibitem[\protect\citeauthoryear{{Sion}, {Solheim}, {Szkody}, {Gaensicke}  \&
  {Howell}}{{Sion} et~al.}{2006}]{Sion2006CPEriHydrogen}
{Sion} E.~M.,  {Solheim} J.-E.,  {Szkody} P.,  {Gaensicke} B.~T.,   {Howell}
  S.~B.,  2006, \mn@doi [\apjl] {10.1086/500085}, \href
  {https://ui.adsabs.harvard.edu/abs/2006ApJ...636L.125S} {636, L125}

\bibitem[\protect\citeauthoryear{{Strohmayer}}{{Strohmayer}}{2005}]{Strohmayer2005timings}
{Strohmayer} T.~E.,  2005, \mn@doi [\apj] {10.1086/430439}, \href
  {https://ui.adsabs.harvard.edu/abs/2005ApJ...627..920S} {627, 920}

\bibitem[\protect\citeauthoryear{{Strohmayer}}{{Strohmayer}}{2008}]{Strohmayer2008HMCncHighResXray}
{Strohmayer} T.~E.,  2008, \mn@doi [\apjl] {10.1086/589439}, \href
  {https://ui.adsabs.harvard.edu/abs/2008ApJ...679L.109S} {679, L109}

\bibitem[\protect\citeauthoryear{{Strohmayer}}{{Strohmayer}}{2021}]{Strohmayer2021HMCnc}
{Strohmayer} T.~E.,  2021, \mn@doi [\apjl] {10.3847/2041-8213/abf3cc}, \href
  {https://ui.adsabs.harvard.edu/abs/2021ApJ...912L...8S} {912, L8}

\bibitem[\protect\citeauthoryear{{Verbunt} \& {Rappaport}}{{Verbunt} \&
  {Rappaport}}{1988}]{Verbunt1988rh}
{Verbunt} F.,  {Rappaport} S.,  1988, \mn@doi [\apj] {10.1086/166645}, \href
  {https://ui.adsabs.harvard.edu/abs/1988ApJ...332..193V} {332, 193}

\bibitem[\protect\citeauthoryear{Virtanen et~al.,}{Virtanen
  et~al.}{2020}]{2020SciPy-NMeth}
Virtanen P.,  et~al., 2020, \mn@doi [Nature Methods]
  {10.1038/s41592-019-0686-2}, \href {https://rdcu.be/b08Wh} {17, 261}

\bibitem[\protect\citeauthoryear{{Webbink}}{{Webbink}}{1984}]{Webbink1984DWDprogenitorsRCrB}
{Webbink} R.~F.,  1984, \mn@doi [\apj] {10.1086/161701}, \href
  {https://ui.adsabs.harvard.edu/abs/1984ApJ...277..355W} {277, 355}

\bibitem[\protect\citeauthoryear{{Wong} \& {Bildsten}}{{Wong} \&
  {Bildsten}}{2021}]{Wong2021AMCVnMESAPureHe}
{Wong} T. L.~S.,  {Bildsten} L.,  2021, \mn@doi [\apj]
  {10.3847/1538-4357/ac2b2a}, \href
  {https://ui.adsabs.harvard.edu/abs/2021ApJ...923..125W} {923, 125}

\bibitem[\protect\citeauthoryear{{Wu}, {Cropper}, {Ramsay}  \&
  {Sekiguchi}}{{Wu} et~al.}{2002}]{Wu2002UImodel}
{Wu} K.,  {Cropper} M.,  {Ramsay} G.,   {Sekiguchi} K.,  2002, \mn@doi [\mnras]
  {10.1046/j.1365-8711.2002.05190.x}, \href
  {https://ui.adsabs.harvard.edu/abs/2002MNRAS.331..221W} {331, 221}

\bibitem[\protect\citeauthoryear{{Ye} et~al.,}{{Ye}
  et~al.}{2019}]{TianQinOptimisingOrbits}
{Ye} B.-B.,  et~al., 2019, \mn@doi [International Journal of Modern Physics D]
  {10.1142/S0218271819501219}, \href
  {https://ui.adsabs.harvard.edu/abs/2019IJMPD..2850121Y} {28, 1950121}

\bibitem[\protect\citeauthoryear{{Zhang} \& {Jeffery}}{{Zhang} \&
  {Jeffery}}{2012}]{Zhang2012FormationOfHeRichHotSubdwarfs}
{Zhang} X.,  {Jeffery} C.~S.,  2012, \mn@doi [\mnras]
  {10.1111/j.1365-2966.2011.19711.x}, \href
  {https://ui.adsabs.harvard.edu/abs/2012MNRAS.419..452Z} {419, 452}

\makeatother
\end{thebibliography}



\newpage
\twocolumn
\appendix

\section{Proper motion deduction}
\label{appendix:propermotionGraphs}
\begin{figure}
    \centering
    \includegraphics[trim={0.5cm 0 0.5cm 0.5cm},clip,width=\columnwidth,keepaspectratio]{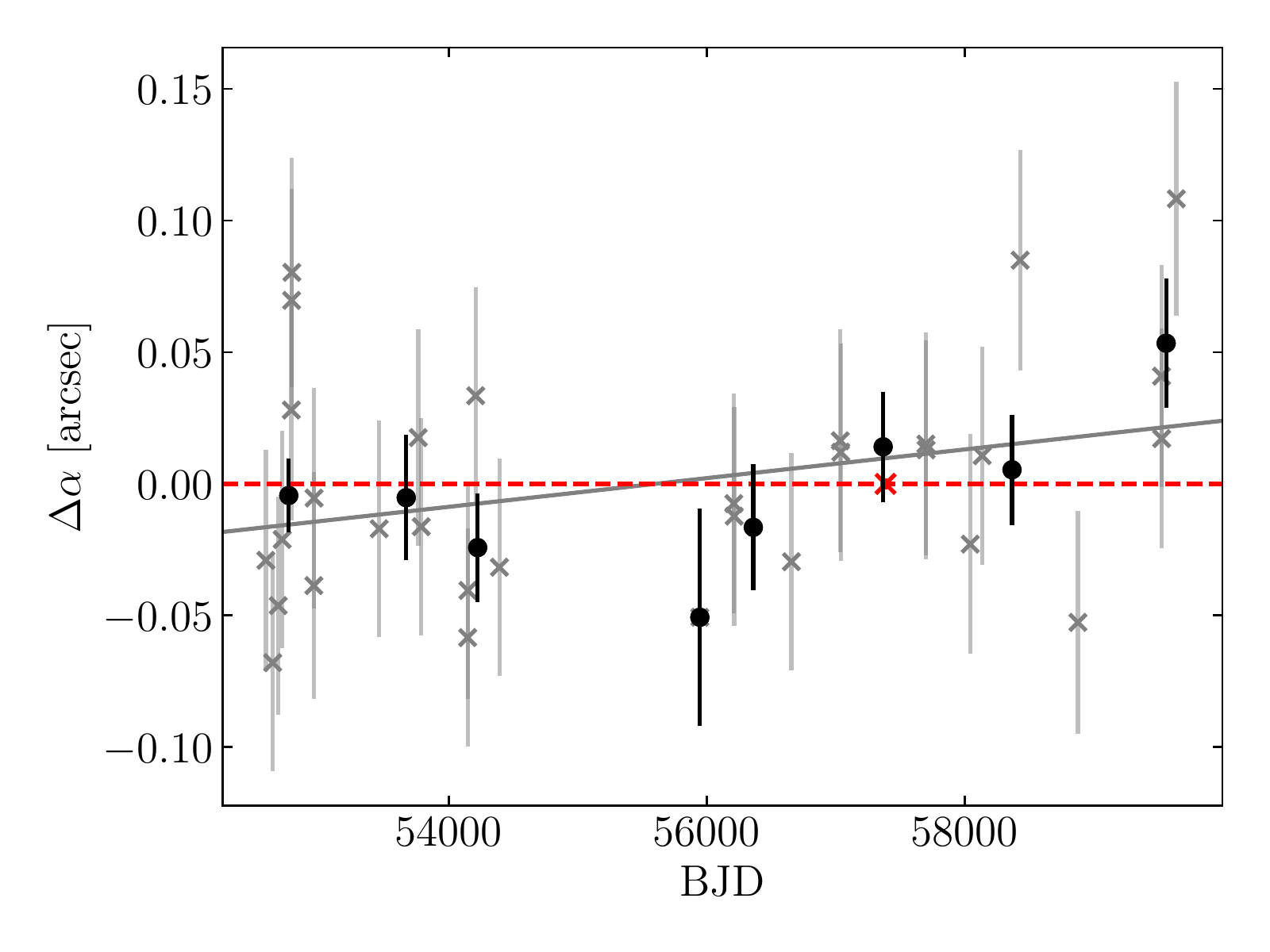}
    \includegraphics[trim={0.5cm 0.5cm 0.5cm 0.cm},clip,width=\columnwidth,keepaspectratio]{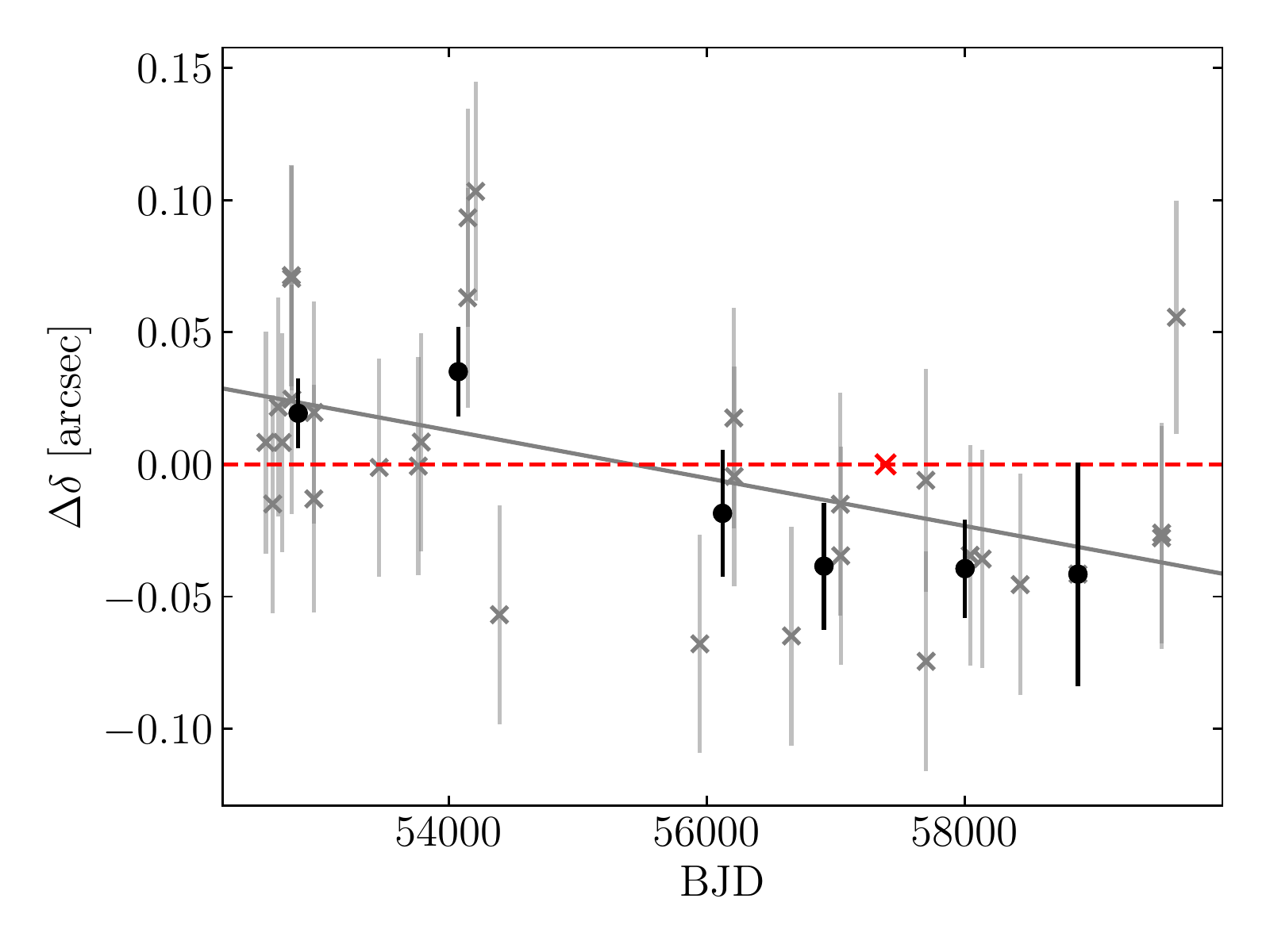}
    \caption{Individual measurements of the right ascension ($\alpha$, top) and declination ($\delta$, bottom), displayed as light grey crosses. Black circles are binned measurements in one thousand day intervals, starting at BJD=52500. The best-fit proper motions are displayed as a solid grey line, while we also constrain upper proper motion limits in \cref{subsec:ProperMotion}. An artificial error has been added to the $\alpha$/$\delta$ of all nights in quadrature, such that the reduced $\chi^2$ of the best-fit proper motions are equal to one. The single red point on each plot marks the coordinate of HM~Cnc stated in Gaia~DR3 (with uncertainty plotted). These Gaia data points are not included in the proper motion fits since they do not include any impact due to atmospheric refraction. The majority of observations are from ULTRACAM, which has a pixel scale of 0.3\arcsec per pixel on all telescopes.}
    \label{fig:my_label}
\end{figure}

\section{Ephemeris corrections}
\label{secappendix:ephemerisCorr}
Small corrections can be made to the obtained frequency derivatives obtained in this study. First by considering the Shklovskii effect \citep[][]{Shklovskii1970effect}, which can be computed with
\begin{equation}
    S_h=V_\perp^2/ (c\,D)
\end{equation}
with $c$ the speed of light and $D$ the distance. With our approximate transverse velocity of 30\,km\,s$^{-1}$ and distance of $1.7\pm0.5$\,kpc from \cref{subsec:ProperMotion},
\begin{equation}
    \dot f_{\textrm{Shk}} = f_0\,S_h = (1.8\pm0.5)\times10^{-22}\,\textrm{Hz\,s}^{-1}
\end{equation}
\begin{equation}
    \ddot f_{\textrm{Shk}}~\approx~\dot f_0\,S_h = (2.0\pm0.6)\times10^{-35}\,\textrm{Hz\,s}^{-2}
\end{equation}
In addition, one can account for acceleration of the system due to the galactic potential, $a_{\textrm{gal}}$, as well. We calculate the following galactic corrections to $\dot f$ and $\ddot f$
\begin{equation}
    \dot f_{\textrm{gal}} = f_0\,a_{\textrm{gal}}/c = 8.3\times10^{-22}\,\textrm{Hz\,s}^{-1}
\end{equation}
\begin{equation}
    \ddot f_{\textrm{gal}} \approx \dot f_0\,a_{\textrm{gal}}/c = 9.5\times10^{-35}\,\textrm{Hz\,s}^{-2}
\end{equation}
\noindent where $a_{\textrm{gal}}$ was computed using the \textsc{MilkyWayPotential} routine of \textsc{gala} \citep[][]{gala} with HM~Cnc placed at a galactic coordinate 1.7\,kpc away from the Sun towards the north galactic pole (in Cartesian coordinates, [-8, 0, 1.7]\,kpc).

These components can then be subtracted from our measured values, hence the corrected $\dot f$ and $\ddot f$ are
\begin{equation}
    \dot f_{\textrm{corr}} = \dot f_0 - \dot f_{\textrm{shk}} - \dot f_{\textrm{gal}}
\end{equation}
\begin{equation}
    \ddot f_{\textrm{corr}} = \ddot f_0 - \ddot f_{\textrm{shk}} - \ddot f_{\textrm{gal}}
\end{equation}
As evident by the magnitude of each correction, all are negligible compared to the uncertainties quoted in Table~\ref{tab:HMCnc_solution}. Increasing the distance to $D=(11\pm3)$\,kpc for halo-like transverse velocities (\cref{subsec:ProperMotion}) results in larger corrections than those stated above. However, neither the correction of $\dot f_0$ and $\ddot f_0$ due to the Shklovskii effect nor galactic acceleration increase by more than $10\times$ and are still negligible compared to the precision of our ephemerides.



\section{Timing solutions and observing log}
\begin{table*}
    \centering
    \caption{The full list of optical timing solutions addressed in this study. All solutions are raw extracted values for each individual filter; the phasing corrections of Table~\ref{tab:offsets} and the 0.014 cycle `flickering error' discussed in \cref{subsec:Flickering} were not applied to these presented values. A clear filter indicates that non-filtered white light was incident on the detector. In the instrument column, UCAM, USPEC and HCAM are abbreviations for ULTRACAM, ULTRASPEC and HiPERCAM.}
    \begin{tabular}{c c c c c}
        \hline
         Time (BJD, TDB, UTC)  & $\Delta$Time (BJD, TDB, UTC) & Telescope & Instrument & Filter \\
         \hline
         51634.033908 & 0.000067 & ESO 3.6m & EFOSC & B\\
51911.047491 & 0.000026 & TNG & DoLoReS & B\\
51911.103190 & 0.000052 & TNG & DoLoReS & B\\ 
52225.293041 & 0.000032 & VLT & FORS & R\\
52580.316805 & 0.000020 & VLT & FORS & R\\
52583.230663 & 0.000023 & TNG & DoLoReS & B\\
52619.187101 & 0.000034 & VLT & FORS & R\\
52636.193912 & 0.000025 & TNG & DoLoReS & V\\
52637.116848 & 0.000017 & TNG & DoLoReS & B\\
52645.084430 & 0.000015 & NOT & ALFOSC & Clear\\
52646.055759 & 0.000012 & NOT & ALFOSC & Clear\\
52647.090283 & 0.000008 & NOT & ALFOSC & Clear\\
52647.142372 & 0.000022 & TNG & DoLoReS & B\\
52675.938780 & 0.000055 & NOT & ALFOSC & Clear\\
52677.036455 & 0.000020 & NOT & ALFOSC & Clear\\ 
52679.035021 & 0.000050 & TNG & DoLoReS & B\\
52709.937606 & 0.000038 & TNG & DoLoReS & B\\
52710.019443 & 0.000075 & TNG & DoLoReS & V\\
52738.901426 & 0.000038 & TNG & DoLoReS & B\\
52760.905962 & 0.000021 & TNG & DoLoReS & B\\ 
52779.907568 & 0.000067 & WHT & UCAM & g'\\
52780.897276 & 0.000072 & WHT & UCAM & i'\\
52780.901042 & 0.000021 & WHT & UCAM & g'\\
52780.901073 & 0.000051 & WHT & UCAM & u'\\
52781.894725 & 0.000060 & WHT & UCAM & u'\\
52781.894733 & 0.000084 & WHT & UCAM & i'\\
52781.894763 & 0.000025 & WHT & UCAM & g'\\
52782.892068 & 0.000062 & WHT & UCAM & u'\\
52782.892226 & 0.000090 & WHT & UCAM & i'\\
52782.899519 & 0.000034 & WHT & UCAM & g'\\
52784.883031 & 0.000099 & WHT & UCAM & u'\\
52784.883042 & 0.000034 & WHT & UCAM & g'\\
52784.883089 & 0.000113 & WHT & UCAM & i'\\
52955.260171 & 0.000070 & WHT & UCAM & u'\\
52955.260253 & 0.000057 & WHT & UCAM & g'\\
52955.260287 & 0.000058 & WHT & UCAM & i'\\
52957.191561 & 0.000039 & WHT & UCAM & g'\\
52957.191619 & 0.000075 & WHT & UCAM & u'\\
52957.195317 & 0.000112 & WHT & UCAM & i'\\
52973.238365 & 0.000018 & INT & WFC & Clear\\
53023.123751 & 0.000029 & NOT & ALFOSC & Clear\\
53024.147096 & 0.000046 & NOT & ALFOSC & Clear\\
53053.080944 & 0.000027 & NOT & ALFOSC & Clear\\ 
53142.904373 & 0.000035 & TNG & DoLoReS & B\\
53461.954123 & 0.000023 & TNG & DoLoReS & V\\ 
53701.264987 & 0.000008 & VLT & UCAM & g'\\
53701.264993 & 0.000016 & VLT & UCAM & r'\\
53701.265027 & 0.000023 & VLT & UCAM & u'\\
53702.232700 & 0.000027 & VLT & UCAM & u'\\
53702.240109 & 0.000009 & VLT & UCAM & g'\\
53702.243817 & 0.000016 & VLT & UCAM & r'\\
53702.288476 & 0.000020 & VLT & UCAM & u'\\
53702.292164 & 0.000014 & VLT & UCAM & r'\\
53702.292206 & 0.000007 & VLT & UCAM & g'\\
53702.329385 & 0.000013 & VLT & UCAM & r'\\
53702.329398 & 0.000008 & VLT & UCAM & g'\\
53702.329419 & 0.000023 & VLT & UCAM & u'\\
53702.359121 & 0.000016 & VLT & UCAM & r'\\
53702.359124 & 0.000011 & VLT & UCAM & g'\\
53702.359171 & 0.000034 & VLT & UCAM & u'\\
53703.244843 & 0.000026 & VLT & UCAM & r'\\
53703.244848 & 0.000013 & VLT & UCAM & g'\\
53703.244881 & 0.000062 & VLT & UCAM & u'\\
\hline
\end{tabular}
    \label{tableappendix:timingSolutionsAll}
\end{table*}
\begin{table*}
    \centering
    \begin{tabular}{c c c c c}
        \hline
         Time (BJD, TDB, UTC)  & $\Delta$Time (BJD, TDB, UTC) & Telescope & Instrument & Filter \\
         \hline

53703.267148 & 0.000016 & VLT & UCAM & r'\\
53703.267165 & 0.000009 & VLT & UCAM & g'\\
53703.267165 & 0.000024 & VLT & UCAM & u'\\
53703.296927 & 0.000013 & VLT & UCAM & r'\\
53703.296960 & 0.000021 & VLT & UCAM & u'\\
53703.300648 & 0.000008 & VLT & UCAM & g'\\
53703.334140 & 0.000016 & VLT & UCAM & r'\\
53703.334146 & 0.000008 & VLT & UCAM & g'\\
53703.334180 & 0.000032 & VLT & UCAM & u'\\
53703.356451 & 0.000013 & VLT & UCAM & r'\\
53703.356470 & 0.000008 & VLT & UCAM & g'\\
53703.360202 & 0.000024 & VLT & UCAM & u'\\
53764.997223 & 0.000025 & TNG & DoLoReS & V\\
53765.983439 & 0.000018 & TNG & DoLoReS & B\\
53788.136902 & 0.000030 & TNG & DoLoReS & V\\
53797.015987 & 0.000019 & TNG & DoLoReS & B\\
54149.909394 & 0.000040 & TNG & DoLoReS & V\\
54149.957790 & 0.000044 & TNG & DoLoReS & B\\
54209.968478 & 0.000032 & TNG & DoLoReS & V\\ 
54394.235140 & 0.000012 & WHT & UCAM & g'\\
54394.235166 & 0.000023 & WHT & UCAM & r'\\
54394.235184 & 0.000029 & WHT & UCAM & u'\\
55532.296391 & 0.000031 & NTT & UCAM & u'\\
55532.300001 & 0.000040 & NTT & UCAM & i'\\
55532.303745 & 0.000011 & NTT & UCAM & g'\\
55947.917226 & 0.000026 & WHT & UCAM & r'\\
55947.920928 & 0.000013 & WHT & UCAM & g'\\
55947.920960 & 0.000027 & WHT & UCAM & u'\\
56211.250500 & 0.000048 & WHT & UCAM & u'\\
56211.250541 & 0.000027 & WHT & UCAM & g'\\
56211.250555 & 0.000049 & WHT & UCAM & r'\\
56214.168047 & 0.000047 & WHT & UCAM & u'\\
56214.168116 & 0.000019 & WHT & UCAM & g'\\
56214.168116 & 0.000034 & WHT & UCAM & r'\\
56342.883367 & 0.000113 & INT & WFC & Clear\\ 
56658.071653 & 0.000030 & WHT & UCAM & r'\\
56658.071729 & 0.000028 & WHT & UCAM & u'\\
56658.075349 & 0.000018 & WHT & UCAM & g'\\
57037.085934 & 0.000034 & WHT & UCAM & u'\\
57037.089626 & 0.000017 & WHT & UCAM & g'\\
57040.181994 & 0.000026 & WHT & UCAM & u'\\
57040.185668 & 0.000013 & WHT & UCAM & g'\\
57040.185680 & 0.000024 & WHT & UCAM & r'\\
57700.349106 & 0.000025 & NTT & UCAM & g'\\
57700.349118 & 0.000049 & NTT & UCAM & r'\\
57700.352702 & 0.000089 & NTT & UCAM & u'\\
57702.347290 & 0.000019 & NTT & UCAM & g'\\
57702.351091 & 0.000059 & NTT & UCAM & u'\\
57702.354797 & 0.000039 & NTT & UCAM & r'\\
58044.245911 & 0.000064 & WHT & HCAM & i$_s$\\
58044.246005 & 0.000045 & WHT & HCAM & r$_s$\\
58044.246034 & 0.000021 & WHT & HCAM & g$_s$\\
58137.164798 & 0.000040 & NTT & UCAM & u$_s$\\
58137.183422 & 0.000026 & NTT & UCAM & r$_s$\\
58137.190855 & 0.000014 & NTT & UCAM & g$_s$\\
58431.326300 & 0.000015 & NTT & UCAM & g$_s$\\
58431.326373 & 0.000058 & NTT & UCAM & i$_s$\\
58431.330021 & 0.000040 & NTT & UCAM & u$_s$\\
58494.047163 & 0.000015 & GTC & HCAM & i$_s$\\
58494.047164 & 0.000008 & GTC & HCAM & r$_s$\\
58494.047166 & 0.000006 & GTC & HCAM & g$_s$\\
58494.047172 & 0.000039 & GTC & HCAM & z$_s$\\
58494.047174 & 0.000015 & GTC & HCAM & u$_s$\\
58879.239177 & 0.000051 & NTT & UCAM & u$_s$\\
58879.246677 & 0.000019 & NTT & UCAM & g$_s$\\
58879.250316 & 0.000078 & NTT & UCAM & i$_s$\\
58880.746361 & 0.000037 & TNO & USPEC & KG5\\
\hline
\end{tabular}
    \label{tableappendix:timingSolutionsAllcont}
\end{table*}
\begin{table*}
    \centering
    \begin{tabular}{c c c c c}
        \hline
         Time (BJD, TDB, UTC)  & $\Delta$Time (BJD, TDB, UTC) & Telescope & Instrument & Filter \\
         \hline 

59259.711306 & 0.000032 & TNO & USPEC & KG5\\
59281.643978 & 0.000046 & TNO & USPEC & KG5\\
59526.333033 & 0.000020 & NTT & UCAM & g$_s$\\
59526.336784 & 0.000050 & NTT & UCAM & u$_s$\\
59527.319131 & 0.000044 & NTT & UCAM & u$_s$\\
59527.319145 & 0.000017 & NTT & UCAM & g$_s$\\
59527.319175 & 0.000057 & NTT & UCAM & i$_s$\\
59528.346290 & 0.000045 & NTT & UCAM & g$_s$\\
59528.346293 & 0.000075 & NTT & UCAM & u$_s$\\
59528.346385 & 0.000079 & NTT & UCAM & i$_s$\\
59642.131890 & 0.000019 & NTT & UCAM & g$_s$\\
59642.131937 & 0.000046 & NTT & UCAM & u$_s$\\
59642.132033 & 0.000102 & NTT & UCAM & i$_s$\\
\hline
    \end{tabular}
    \label{tableappendix:timingSolutionsAllcont2}
\end{table*}

\begin{table*}
    \label{tableappendix:ObservingLog}
    \caption{An observing log of all HM~Cnc observations acquired by us with HiPERCAM, ULTRACAM or ULTRSASPEC. Details of observations with archival data can be found at the appropriate studies cited in \cref{subsec:photometry}. Included in this table are the conditions for the data that we obtained. Unless mentioned in the comments, a night is not noticeably impacted by clouds. The duration represents the time that the telescope was on target after acquisition. MJD$_{\textrm{mid}}$ is the MJD at the centre of the observing period. In the instrument column, UCAM, USPEC and HCAM are abbreviations for ULTRACAM, ULTRASPEC and HiPERCAM. Asterisked comments indicate data that were presented in \citet[][]{Barros2007HMCnc}, though are reanalysed.}
    \centering
    \begin{tabular}{c c c c c c c c}
        \hline
         Night & MJD$_\textrm{mid}$  & Filters & Telescope & Instrument & Cadence (s) & Duration (min) & Comments\\
         \hline
2003-05-20  &  52779.9  &  u'~g'~i'  &  WHT  &  UCAM   &  20.0  &  20.3  &  Seeing 1.5-2.0''  \\
2003-05-21  &  52780.9  &  u'~g'~i'  &  WHT  &  UCAM   &  9.8  &  53.2  &  *Seeing 1.2''  \\
2003-05-22  &  52781.9  &  u'~g'~i'  &  WHT  &  UCAM   &  9.8  &  57.9  &  *Seeing 1.0''  \\
2003-05-23  &  52782.9  &  u'~g'~i'  &  WHT  &  UCAM   &  9.8  &  62.4  &  *Seeing 1.0''  \\
2003-05-25  &  52784.9  &  u'~g'~i'  &  WHT  &  UCAM   &  9.8  &  41.2  &  *Seeing 1.3''  \\
2003-11-11  &  52955.2  &  u'~g'~i'  &  WHT  &  UCAM   &  10.0  &  66.1  &  Seeing 1.0''  \\
2003-11-13  &  52957.1  &  u'~g'~i'  &  WHT  &  UCAM   &  10.0  &  35.5  &  Seeing 1.0-1.3''  \\
2005-11-26  &  53701.2  &  u'~g'~r'  &  VLT  &  UCAM   &  3.0  &  107.5  &  *Initially poor seeing, later 1.0-1.5''  \\
2005-11-27  &  53702.3  &  u'~g'~r'  &  VLT  &  UCAM   &  1.5  &  215.5  &  *Seeing < 1.0''  \\
2005-11-28  &  53703.3  &  u'~g'~r'  &  VLT  &  UCAM   &  1.5  &  195.9  &  *Seeing < 1.0''  \\
2007-10-20  &  54394.2  &  u'~g'~r'  &  WHT  &  UCAM   &  9.7  &  102.2  &  Seeing 1.0--1.2''  \\
2010-12-01  &  55532.3  &  u'~g'~i'  &  NTT  &  UCAM   &  19.7  &  193.6  &  Seeing 1.0-1.5''. Intermittent clouds  \\
2012-01-21  &  55947.9  &  u'~g'~r'  &  WHT  &  UCAM   &  10.0  &  113.3  &  Seeing 1.0-1.5''  \\
2012-10-10  &  56211.2  &  u'~g'~r'  &  WHT  &  UCAM   &  8.0  &  32.2  &  Seeing 1.0''  \\
2012-10-13  &  56214.1  &  u'~g'~r'  &  WHT  &  UCAM   &  8.5  &  49.9  &  Seeing 1.4''  \\
2013-12-31  &  56658.0  &  u'~g'~r'  &  WHT  &  UCAM   &  10.0  &  57.2  &  Seeing 1.5''  \\
2015-01-14  &  57037.1  &  u'~g'~i'  &  WHT  &  UCAM   &  10.0  &  66.0  &  Flares in seeing; 1.0--3.0''  \\
2015-01-17  &  57040.2  &  u'~g'~r'  &  WHT  &  UCAM   &  10.0  &  107.2  &  Seeing 1.0--1.4''  \\
2016-11-07  &  57700.3  &  u'~g'~r'  &  NTT  &  UCAM   &  10.0  &  35.4  &  Seeing 1.5''  \\
2016-11-09  &  57702.3  &  u'~g'~r'  &  NTT  &  UCAM   &  10.0  &  56.9  &  Seeing 1.2''  \\
2018-01-18  &  58137.2  &  u$_s$~g$_s$~r$_s$  &  NTT  &  UCAM   &  20.0  &  141.1  &  Seeing 1.5''  \\
2018-11-08  &  58431.3  &  u$_s$~g$_s$~i$_s$  &  NTT  &  UCAM   &  10.0  &  84.7  &  Seeing 1.3''  \\
2020-01-30  &  58879.2  &  u$_s$~g$_s$~i$_s$  &  NTT  &  UCAM   &  10.0  &  147.8  &  Seeing 1.5''  \\
2021-11-07  &  59526.3  &  u$_s$~g$_s$~i$_s$  &  NTT  &  UCAM   &  10.0  &  74.6  &  Focus issues. Good conditions.  \\
2021-11-08  &  59527.3  &  u$_s$~g$_s$~i$_s$  &  NTT  &  UCAM   &  10.0  &  129.7  &  Seeing 1.0--1.2''  \\
2021-11-09  &  59528.3  &  u$_s$~g$_s$~i$_s$  &  NTT  &  UCAM   &  10.0  &  20.4  &  Seeing 1.1''  \\
2022-03-03  &  59642.1  &  u$_s$~g$_s$~i$_s$  &  NTT  &  UCAM   &  10.0  &  71.4  &  Seeing 1.2''  \\
2017-10-17  &  58044.2  &  g$_s$~r$_s$~i$_s$~z$_s$  &  WHT  &  HCAM  &  13.9  &  24.8  &  Seeing 1.2''  \\
2019-01-10  &  58494.0  &  u$_s$~g$_s$~r$_s$~i$_s$~z$_s$  &  GTC  &  HCAM  &  10.0  &  43.3  &  Seeing 0.9''  \\
2020-02-01  &  58880.7  &  KG5  &  TNO  &  USPEC  &  15.4  &  160.2  &  Seeing 1.5''  \\
2021-02-14 & 59259.7&KG5&TNO&USPEC&14.2&96.7&Seeing 1.0--1.4''\\
2021-03-08  &  59281.6  &  KG5  &  TNO  &  USPEC  &  16.4  &  175.4  &  Seeing 0.9''  \\
         \hline
    \end{tabular}
    \label{tab:observingLog}
\end{table*}

\section{MESA input models}
\begin{table*}
   \centering
   \caption{The initial parameters of the MESA models presented in Fig.~\ref{fig:MESAresults}. The donor mass stated includes the mass of the diffused hydrogen envelope. The present day system mass is not the combination of the starting masses due to the mass loss mechanisms invoked, as discussed in \cref{subsec:BinaryModelling}.}
   \begin{tabular}{c c c c}
        \hline
        Donor Mass (\(\textup{M}_\odot\)) & Accretor Mass (\(\textup{M}_\odot\)) & Hydrogen Envelope (10$^{-3}$ \(\textup{M}_\odot\)) & Fig.~\ref{fig:MESAresults} Label\\
        \hline
        0.16 & 1.20 & 0.1 & 0.156\,\(\textup{M}_\odot\) + 1.202\,\(\textup{M}_\odot\)\\
        0.16 & 1.30 & 0.1 & 0.156\,\(\textup{M}_\odot\) + 1.302\,\(\textup{M}_\odot\)\\
        0.17 & 1.00 & 0.1 & 0.167\,\(\textup{M}_\odot\) + 1.001\,\(\textup{M}_\odot\)\\
        0.16 & 1.20 & 1.0 & 0.156\,\(\textup{M}_\odot\) + 1.202\,\(\textup{M}_\odot\)\\
        0.17 & 1.00 & 1.0 & 0.166\,\(\textup{M}_\odot\) + 1.002\,\(\textup{M}_\odot\)\\
        0.17 & 1.20 & 3.0 & 0.162\,\(\textup{M}_\odot\) + 1.205\,\(\textup{M}_\odot\)\\
        0.17 & 1.30 & 3.0 & 0.162\,\(\textup{M}_\odot\) + 1.305\,\(\textup{M}_\odot\)\\
        0.18 & 0.95 & 3.0 & 0.172\,\(\textup{M}_\odot\) + 0.955\,\(\textup{M}_\odot\)\\
        0.176 & 1.25 & 5.6 & 0.163\,\(\textup{M}_\odot\) + 1.258\,\(\textup{M}_\odot\)\\
        0.179 & 1.15 & 5.3 & 0.169\,\(\textup{M}_\odot\) + 1.156\,\(\textup{M}_\odot\)\\
        0.195 & 0.85 & 4.9 & 0.183\,\(\textup{M}_\odot\) + 0.856\,\(\textup{M}_\odot\)\\
        \hline
   \end{tabular}
   \label{tab:my_label2}
\end{table*}


\bsp	
\label{lastpage}
\end{document}